\documentclass[a4paper]{article}
\usepackage[utf8]{inputenc}
\usepackage[T1]{fontenc}
\usepackage{fullpage}
\usepackage{comment}
\usepackage{amssymb}
\usepackage{hyperref} 
\usepackage{amsmath,bm,color}
\usepackage{amsthm}
\usepackage{mathtools}
\usepackage{graphicx}
\usepackage{subcaption}
\usepackage{authblk}
\usepackage{upref}
\usepackage[style=apa,natbib=true,backend=biber]{biblatex}
\renewcommand{\cite}{\textcite}
\definecolor{purple}{rgb}{.4,0,.4}
\definecolor{grey}{rgb}{0.5, 0.50, 0.5}

\DeclareMathOperator{\logit}{logit}
\DeclareMathOperator{\Var}{Var}
\DeclareMathOperator{\Cov}{Cov}

\DeclareMathOperator{\MAF}{MAF}
\DeclareMathOperator{\sgn}{sgn}

\newtheorem{theorem}{Theorem}

\newtheorem{observation}[theorem]{Observation}

\AtBeginDocument{ 
  \label{CorrectFirstPageLabel} 
} 

\author[1,2]{Pål Vegard Johnsen}
\author[2]{Øyvind Bakke}
\author[2]{Thea Bjørnland}
\author[3]{Andrew Thomas DeWan}
\author[2]{Mette Langaas}

\affil[1]{SINTEF Digital, Oslo, Norway}
\affil[2]{Department of Mathematical Sciences, Norwegian University of Science and Technology, Trondheim, Norway}
\affil[3]{Department of Chronic Disease Epidemiology and Center for Perinatal, Pediatric and Environmental Epidemiology, Yale School of Public Health}

\title{Saddlepoint approximations in binary genome-wide association studies}

\date{}

\begin{document}
\maketitle

\begin{abstract}
    We investigate saddlepoint approximations applied to the score test statistic in genome-wide association studies with binary phenotypes. The inaccuracy in the normal approximation of the score test statistic increases with increasing sample imbalance and with decreasing minor allele count. Applying saddlepoint approximations to the score test statistic distribution greatly improve the accuracy, even far out in the tail of the distribution. By using exact results for an intercept model and binary covariate model, as well as simulations for models with nuisance parameters, we emphasize the need for continuity corrections in order to achieve valid $p$-values. The performance of the saddlepoint approximations is evaluated by overall and conditional type I error rate on simulated data. We investigate the methods further by using data from UK Biobank with skin and soft tissue infections as phenotype, using both common and rare variants. The analysis confirms that continuity correction is important particularly for rare variants, and that the normal approximation gives a highly inflated type I error rate for case imbalance.
\end{abstract}

\section{Introduction}
We consider score tests for logistic regression models in which the response is imbalanced and the covariate of interest is discrete and skewed. This typically occurs in a genome-wide association study (GWAS) with binary phenotypes, henceforth denoted binary GWAS, where one of the phenotypes is rare.

In a GWAS each single nucleotide polymorphism (SNP) is tested individually for association with a particular phenotype. In a modern biobank including several hundred thousands SNPs, rejection of the null hypothesis needs to be evaluated with a very low $p$-value threshold, typically equal to $5 \cdot 10^{-8}$, in order to control the family-wise error rate (FWER). In a binary GWAS with imbalanced response, new challenges arise.

As an example, we consider a follow-up study on skin and soft tissue infection (SSTI) using UK-biobank data, motivated by \cite{rogne_gwas_2021}. Using data on unrelated white Caucasians with no prior history of SSTI at recruitment, we obtain 6.5 years of follow-up data on approximately $300 \, 000$ individuals, out of which approximately $0.7\%$ where diagnosed with SSTI during follow-up, and classified as cases. The overall sample size may be large, but if there are few cases or controls with a certain genotype, relying on asymptotic normality of the score test statistic may yield spurious results. In fact, the score test applied under asymptotic theory yields invalid $p$-values if the case proportion is too small. In addition, the severity in this flaw increases with decreasing minor allele frequencies (MAF). Both \cite{ma_recommended_2013} and \cite{dey_fast_2017} have illustrated this issue for sample sizes of up to $20 \, 000$ individuals of which between $1 \%$ and $10 \%$ were cases. Motivated by the UK-Biobank SSTI data set, we show that the normal approximation can be flawed even when the total sample size is in the order of several hundred thousands.  A solution proposed by \cite{ma_recommended_2013} is to apply the \cite{firth1993bias} bias-corrected logistic regression test. The test gives valid $p$-values when the imbalance is not too severe, and it is at the same time less conservative than the likelihood ratio test. As Firth's test is computationally inefficient for genome-wide testing, a test based on a saddlepoint approximation to the score statistic was proposed by \cite{dey_fast_2017}. This so-called SPA-test showed good properties yielding both valid or close to valid $p$-values even when Firth's test failed to do so, as well as being as powerful as Firth's test. 

Our theoretical contribution to the ongoing development of valid score tests for genome-wide association studies with imbalanced binary phenotypes is twofold. First, we establish the discrete and bounded nature of the score, and derive the exact conditional distribution of the score test statistic for two particular examples of logistic regression models, namely models with intercept and genetic variant only, as well a models with an additional binary nuisance covariate. Second, we propose continuity-corrected saddlepoint approximations to the conditional distribution of the score statistic. We compare our proposed method against exact results as well as the approach introduced in \cite{dey_fast_2017}. We study the validity of tests both conditionally and unconditionally. 

We show that a score test derived from the efficient score, or equivalently a null-orthogonal reparameterization of the logistic regression model, coincides with the SPA-test by \cite{dey_fast_2017}, thus providing a novel interpretation of the SPA-test as a two-step approximation to the conditional distribution of the score statistic.

We study our proposed continuity-corrected saddlepoint approximations as well as other existing methods, using the follow-up study of SSTIs as explained above, and on simulated data. 

\section{The score test statistic for logistic regression models in GWAS}

\subsection{Notation, statistical model and hypotheses}\label{notstathyp}

We consider tests for genotype--phenotype associations in large cohorts or populations. We assume that binary phenotypes, $Y_i$, non-genetic covariates $\bm x_i$ and allele counts $g_i$ for a single variant, \ $i=1$, \ldots, $n$, have been collected from $n$ individuals. We consider directly biallelic allele counts in which $g_i \in (0,1,2)$. We model the relationship between the response and the covariates in a logistic regression model in which the $Y_i$ are independent and Bernoulli distributed with success probability $\mu_i$ and
\begin{equation}
 \logit\mu_i=\bm x^\text T_i\bm\beta+\gamma g_i,\label{model}
\end{equation}
$i=1$, \ldots, $n$. Here, $\bm x_i$ is a vector of dimension $d$ containing 1 (corresponding to an intercept) and $d-1$ covariates, $\bm\beta$ a $d$-dimensional vector of nuisance parameters and $\gamma$ the parameter of interest. Our aim is to perform the hypothesis test
\begin{equation}
 H_0\colon\gamma = 0 \quad\text{against}\quad H_1\colon\gamma \neq 0.
    \label{eq:hyp1}
\end{equation}
In a GWAS, the test is performed multiple times, for different genetic variants. To control the FWER at a 5\% level in GWAS involving common variants, a significance level of $5\cdot10^{-8}$ is commonly used for each test \citep{jannot2015p}.

\subsection{The score test statistic}\label{propscore}

The score vector is the gradient of the log-likelihood function with respect to the parameters, which for the logistic regression model \eqref{model} is 
\begin{equation}{
 \bm{U}=\begin{pmatrix}\bm U_{\bm \beta}\\U_\gamma\end{pmatrix} 
  = \begin{pmatrix}X^\text T(\bm Y-\bm\mu)\\\bm g^\text T(\bm Y - \bm\mu)\end{pmatrix}},\label{scoredef}
\end{equation}
where $\bm Y$ and $\bm g$ are column vectors of length $n$ with $Y_i$ and $g_i$ as elements respectively, $\bm\mu=E\bm Y$, and $X$ is an $n \times d$ matrix with $\bm x_i^\text T$ as rows. We have partitioned the score vector according to the parameter of interest, $\gamma$, and the nuisance parameters, $\bm\beta$. The score vector has mean $\bm0$ and covariance matrix, by definition referred to as the expected Fisher information
\begin{equation}
 F= {\begin{pmatrix}F_{\bm \beta \bm \beta}&\bm F_{\gamma\bm \beta}^\text T\\\bm F_{\gamma \bm \beta}&F_{\gamma\gamma}\end{pmatrix}}
  =\begin{pmatrix}X^\text TWX&X^\text TW\bm g\\\bm g^\text TWX& \bm g^\text TW\bm g\end{pmatrix},
\label{fisher}
\end{equation}
where $W$ is a diagonal matrix with $\mu_i(1-\mu_i)$ as the $ii$ entry.

Using the score test, the null hypothesis of~\eqref{eq:hyp1} is rejected if there is sufficient distance between the null value $\gamma = 0$ and the maximum likelihood estimate of~$\gamma$. To judge this distance, without actually calculating the estimate, one uses the partial derivative $U_\gamma$ of the log-likelihood with respect to $\gamma$ at $\gamma = 0$, along with the probability distribution of $U_\gamma$ under the null. The proof of the following observation is given in Appendix~\ref{obsproofs}.

\begin{observation}
When $g_i \in (0,1,2)$, the score $U_{\gamma}$ with respect to $\gamma$ is a bounded lattice random variable with support on $-\bm g^{\textup T} \bm \mu$, $1 -\bm g^{\textup T} \bm{\mu}$, $2 -\bm g^{\textup T} \bm{\mu}$, \ldots, $\bm g^\textup T\bm1-\bm g^\textup T\bm\mu$. 
\label{obs:discrete_bounded}
\end{observation}

Importantly, the score is -- as in our situation -- often a function of unknown nuisance parameters. Then, one may consider the \emph{conditional} null distribution of the score for the parameter of interest, $U_\gamma$, given that the components of the score vector corresponding to the nuisance parameters are equal to zero, $\bm U_{\bm \beta}=\bm0$ \citep[see e.g.][]{smyth2003pearson}. In this conditional framework, the unknown nuisance parameters are equal to the corresponding maximum likelihood estimates calculated under the null hypothesis $\gamma = 0$, so that $U_\gamma=\bm g^\text T(\bm Y-\hat{\bm\mu})$, where $\hat{\bm\mu}$ consists of the fitted values of the null model. However, this conditional score test statistic will still be a lattice random variable, yet with a narrower support than described in Observation \ref{obs:discrete_bounded}. See Appendix \ref{supportCondScore}.

In many applications, one may approximate the distribution of the score vector $\bm U$ by a multivariate normal distribution with mean $\bm 0$ and covariance matrix $F$. The conditional distribution of $U_{\gamma}$ given $\bm U_{\bm \beta} =\nobreak\bm 0$ under the null ($\gamma=0$) is then asymptotically a normal distribution with mean 0 and variance
\begin{equation}
 \tilde{F}_{\gamma \gamma}(\hat{\bm \mu}) = \bm g^\text T \hat{W} \bm g - \bm g^\text T \hat{W} X (X^\text T \hat{W} X)^{-1} X^\text T \hat{W} \bm g,
 \label{condvar}
\end{equation}
where $\hat{W}$ is the diagonal matrix with $\hat{\mu}_i(1-\hat{\mu}_i)$ as the $ii$ entry. As outlined in the Introduction, the normal approximation to the score vector may lead to spurious results for genotype--phenotype associations when the phenotype is a binary variable. For example, even if the the overall sample size is large, the normal approximation may be inaccurate if the sample contains few individuals with response $y_i = 1$ (e.g., having the disease under study) and genotype $g_i>0$ (carrying the minor allele).

In the next section, we present a score test for \eqref{eq:hyp1} based on a double saddlepoint approximation to the conditional null distribution of the score statistic $U_{\gamma}$ for the logistic regression model \eqref{model}, given $\bm U_{\bm \beta} = \bm 0$. Here, we first state two observations that give the \emph{exact} conditional null distribution for two special cases of the regression model \eqref{model}. Proofs are given in Appendix~\ref{obsproofs}.

\begin{observation}
Consider a logistic regression model as in \eqref{model}, but with $\logit\mu_i = \beta + \gamma g_i$, henceforth denoted the \textit{intercept model}. Let $n_j$ be the number of individuals with genotype $g_i = j$, \ $j = 0, 1, 2$, and let $\logit\mu = \beta$. Then, the null distribution of $U_{\gamma}$ given $U_{\beta} = 0$ is a sum of trivariate hypergeometric point probabilities,
\newlength{\lengde}\settowidth{\lengde}{$\displaystyle\sum_{k=\max(\lceil(u^*-n_1)/2\rceil,0)}^{\min(\lfloor u^*/2\rfloor,n_2)}$}
\newlength{\llengde}\settowidth{\llengde}{$\displaystyle\sum^{\min(\lfloor u^*/2\rfloor,n_2)}$}
\setlength{\lengde}{.5\lengde}\addtolength{\lengde}{.5\llengde}
\[
 P(U_{\gamma} = u \mid U_{\beta} = 0) = \smashoperator[r]{\sum_{(v_0,v_1,v_2)\in S}} \frac{\binom{n_0}{v_0}\binom{n_1}{v_1}\binom{n_2}{v_2}}{\binom{n}{n \mu}}
   =\mathop{\mathmakebox[\lengde][r]{\sum_{\crampedclap{k=\max(\lceil(u^*-n_1)/2\rceil,0)}}^{\cramped{\min(\lfloor u^*/2\rfloor,n_2)}}}}\frac{\binom{n_0}{n\mu -u^* + k}\binom{n_1}{u^*-2k}\binom{n_2}{k}}{\binom{n}{n \mu}},
\]
where the sum is taken over all triples $(v_0, v_1, v_2)$ of integers in the set $S$ defined by $0 \leq v_j \leq n_j$ for $j = 0$, $1$, $2$, \ $v_0 + v_1 + v_2 = n\mu$ and $v_1 + 2v_2 = u^*$, and $u^*=u+(n_1+2n_2)\mu$. The function outputs $\lceil x \rceil$ and $\lfloor x \rfloor$ denote the least integer greater than or equal to $x$ (ceiling), and the largest integer less than or equal to $x$ (floor) respectively. 
\label{obs:exact1}
\end{observation}

\begin{observation}
Consider a logistic regression model as in \eqref{model}, where $\logit\mu_i = \beta_0 + \beta_1 x_i + \gamma g_i$, and~$x_i$ is a binary covariate taking value $0$ or $1$ \textup(model with intercept and one binary non-genetic covariate\textup). Let $l_j$ be the number of individuals with $x_i = 0$ and genotype $g_i = j$, \ $j = 0$, $1$, $2$, and let $l=l_0+l_1+l_2$. Define similar counts $m_j$ and $m$ for individuals with $x_i = 1$. Let $\logit\mu_0 = \beta_0$, and $\logit\mu_1 = \beta_0 + \beta_1$. Then, under the null hypothesis,
\[
    P(U_{\gamma} = u \mid\bm U_{\bm \beta} = \bm 0) = \smashoperator[r]{\sum_{\bm s \in S}} \frac{\binom{l_0}{v_0}\binom{l_1}{v_1}\binom{l_2}{v_2}}{\binom{l}{l \mu_0}} \frac{\binom{m_0}{w_0}\binom{m_1}{w_1}\binom{m_2}{w_2}}{\binom{m}{m \mu_1}},
\]
where the sum is taken over all sextuples $\bm s=(v_0, v_1, v_2, w_0, w_1, w_2)$ of integers in the set $S$ defined by $0 \leq v_j \leq l_j$, \ $0 \leq w_j \leq m_j$ for $j = 0$, $1$, $2$, \ $v_0 + v_1 + v_1 = l \mu_0$, \ $w_0 + w_1 + w_2 = m \mu_1$ and $v_1+2v_2 - (l_1 + 2l_2) \mu_0 + w_1 + 2w_2 - (m_1 + 2m_2) \mu_1 = u$.
\label{obs:exact2}
\end{observation}

From Observations \ref{obs:exact1} and \ref{obs:exact2}, it follows that an \textit{exact} $p$-value for the hypothesis test \eqref{eq:hyp1} can be computed for these two special cases of the logistic regression model \eqref{model}. An extension of Observation \ref{obs:exact2} can also be derived for regression models with more categorical covariates. However, for more complex covariate patterns, this approach becomes computationally infeasible, or even intractable when continuous covariates are included. The next section introduces a method of computing $p$-values using double saddlepoint approximation. 

\section{Double saddlepoint approximation}

Tail probabilities $P(U_{\gamma} \geq u \mid \bm U_{\bm \beta} = \bm 0)$ may be estimated by \emph{double saddlepoint approximation} \citep{butler_2007}. This will require the \emph{cumulant generating function} of $\bm{U}=\smash{\begin{pmatrix}\bm U_{\bm \beta}^\text T&U_\gamma\end{pmatrix}}^\text T
  = \smash{\begin{pmatrix}X&\bm g\end{pmatrix}}^\text T(\bm Y-\bm\mu)$ (Section~\ref{propscore})
and of $\bm U_{\bm \beta}$.

\subsection{Cumulant generating function}

The joint cumulant generating function of $\bm U$ is defined by $K(\bm t)=\ln E\left(e^{\bm t^\text T \bm U}\right)$, were $\bm t$ is a vector of dimension $d+1$. By using the fact that $Y_i$ is Bernoulli distributed with parameter $\mu_i$ (Section~\ref{notstathyp}), we obtain
\begin{align}
K(\bm t) &= \sum_{i=1}^n\Bigl(\ln\bigl(1-\mu_i+\mu_ie^{\bm t^\text T\bm z_i}\bigr)-\mu_i\bm t^\text T \bm z_i\Bigr),\label{K}\\
\nabla K(\bm t) &= \sum_{i=1}^n \mu_i\biggl(\frac{1}{(1-\mu_i)e^{-\bm t^{\text T}\bm z_{i}}+\mu_i}-1\biggr)\bm z_i, \quad\text{and}\\
H(\bm t) &= \sum_{i=1}^n \frac{\mu_i(1-\mu_i)e^{-\bm t^{\text T}\bm z_{i}}}{\bigl((1-\mu_i)e^{-\bm t^{\text T}\bm z_{i}}+\mu_i\bigr)^2}\bm{z}_i\bm{z}_i^\text T,
\label{kdoubleprime}
\end{align}
where $\nabla K$ and $H$ denote the gradient and the Hessian of $K$, respectively, and $\bm z_i=\smash{\begin{pmatrix}\bm x_i^\text T&g_i\end{pmatrix}}^\text T$. The cumulant generating function of $\bm U_{\bm \beta}$, its gradient and Hessian, $K_{\bm \beta}$, $\nabla K_{\bm \beta}$ and $H_{\bm \beta}$, respectively, are obtained by replacing $\bm z_i$ by $\bm x_i$ and letting $\bm t$ have dimension $d$ in \eqref{K}--\eqref{kdoubleprime}.

\subsection{Approximated tail probabilities with continuity correction}\label{dspacc}

The survival function (right-tail probability) $S(u)=P(U_{\gamma} \geq u\mid\bm U_{\bm \beta} = \bm0)$ can be approximated as given by \cite{barndorff-nielsen_approximate_1990},
\begin{equation}
 \hat S(u) = 1- \Phi\Bigl( w - \frac{1}{w} \ln\frac{v}{w}\Bigr), 
\label{eq:BN}
\end{equation}
where $\Phi$ denotes the standard normal cumulative distribution function. To approximate the conditional survival function of a lattice random variable we have chosen the double saddlepoint survival approximation with the so-called second continuity correction. Using $f(\bm t_1,\bm t_2)$ as shorthand for $f\bigl(\smash{\begin{pmatrix}\bm t_1^\text T&\bm t_2^\text T\end{pmatrix}}^\text T\bigr)$, where $f$ is a function and $\bm t_1$, $\bm t_2$ vectors, we have
\[
\begin{split}
w&=\sgn(\hat t_{\gamma})\sqrt{2\biggl(-K(\hat{\bm t}_{\bm \beta},\hat{t}_{\gamma})+ \hat{t}_{\gamma}\Bigl(u-\frac12\Bigr)\biggr)}\quad\text{and}\\
v&=2 \Bigl(\sinh\frac{\hat{t}_{\gamma}}2\Bigr) \sqrt{\frac{\det H(\hat{\bm t}_{\bm \beta},\hat{t}_{\gamma})}{\det H_{\bm \beta}(\bm0)}},
\end{split}
\]
where $\smash{\begin{pmatrix}\hat{\bm t}_{\bm \beta}^\text T&\hat{t}_{\gamma}\end{pmatrix}}^\text T$ is the \emph{saddlepoint} satisfying $\nabla K(\hat{\bm t}_{\bm \beta},\hat{t}_{\gamma}) = \smash{\begin{pmatrix}\bm 0^\text T &u-1/2\end{pmatrix}}^\text T$ (\citeauthor{skovgaard1987saddlepoint}, \citeyear{skovgaard1987saddlepoint}, see \citeauthor{butler_2007}, \citeyear{butler_2007}, p.114). In general, also the $d$-dimensional vector $\tilde{\bm t}_{\bm \beta}$ satisfying $\nabla K_{\bm \beta}(\tilde{\bm t}_{\bm \beta})=\bm0$ is involved in the expressions for $w$ and $v$, but $\tilde{\bm t}_{\bm \beta}=\bm0$ in our case (see Appendix \ref{Sol_t_tilde}). Left-tail probabilities can be approximated, taking into account that $U_\gamma$ is a lattice variable with step~1, by $P(U_{\gamma} \leq u\mid\bm U_{\bm \beta} = \bm0)=1-S(u+1)$.

\subsection{Two-sided \emph p-values}\label{twosidedStrategy}

By assuming the score test statistic to have a normal distribution, and for some observation $u$, a two-sided $p$-value is reasonable and given by $P(|U_{\gamma}| \geq |u| \mid \bm U_{\bm \beta} = \bm 0)$ (under the null). However, As the score test statistic has a lattice distribution, the point $-u$ might not be on the grid. If so, the closest grid point to $-u$ farthest away from zero is obtained by $u_{inv} = u - \sgn(u)\cdot  \lceil 2 \cdot \lvert u \rvert\rceil$. We define a two-sided p-value, assuming $u$ positive, to be $P(U_{\gamma} \geq u \mid \bm U_{\bm \beta} = \bm 0) + P(U_{\gamma} \leq u_{inv}\mid \bm U_{\bm \beta} = \bm 0)$, and vice versa when $u$ is negative. 

\begin{figure}[ht]
\centering
\includegraphics[width=1\textwidth]{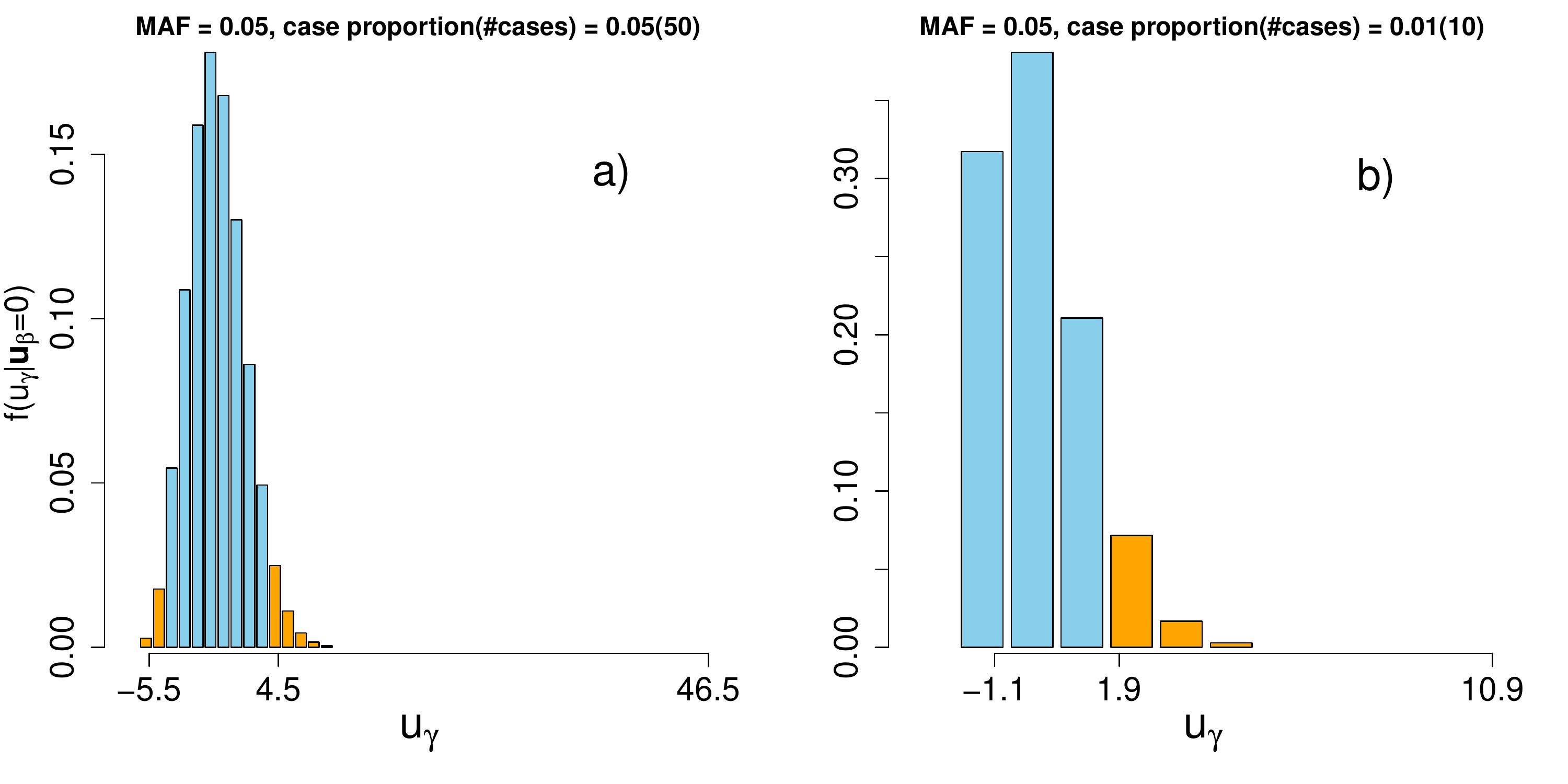}
\caption{The exact lattice distribution of the score test statistic for the intercept model for different case proportions (genotype vector fixed, 1000 individuals). Included is the support $[u^{min},u^{max}]$ of the lattice distribution in each case together with an example of an observed statistic in between, as well as the corresponding computed p-value coloured in orange. The deviation from normal distribution increases for decreasing case proportion. When the distribution is sufficiently skewed, a one-sided p-value is computed.}
\label{DiffInSkewMAF}
\end{figure}

An example is given in Figure \ref{DiffInSkewMAF}a where the exact lattice distribution of the score test statistic under the null hypothesis is given for the intercept model with a genotype vector simulated with $\MAF = 0.05$ and a case proportion of 0.05 ($n = 1000$). Included is the support of the lattice distribution $[u^{min},u^{max}] = [-5.5,46.5]$. An observed $u = 4.5$ will then give $u_{inv} = -4.5$, a situation where $u_{inv} = -u$. The p-value is then equal to the sum of the bars coloured in orange.
The deviation from the normal distribution increases for decreasing case proportion, as can be seen when comparing Figure \ref{DiffInSkewMAF}a to \ref{DiffInSkewMAF}b, where the case proportion is reduced to 0.01 while keeping the same genotype vector in Figure \ref{DiffInSkewMAF}b. In fact, the skewness increases for decreasing case proportion such that the probability mass of the distribution is concentrated on the left, with a longer right tail. Consequently, the score test statistic is asymmetric as well as bounded, which means the point $u_{inv}$ might be outside the support of the lattice distribution. In that case, a one-sided p-value will be computed as seen in Figure 1b with bars coloured orange only to the right of the observed $u = 1.9$ ($u_{inv} = -2.1 < u^{min} = -1.1$). The same observation of increased skewness can be seen with a fixed case proportion, but decreasing MAF.

\section{Single saddlepoint approximation using the efficient score}

Our proposed method is related to the SPA-test by \cite{dey_fast_2017}, which is also based on a saddlepoint approximation to the distribution of a score test statistic. In this section, we provide a novel interpretation of the SPA-test as a two-step approximation to conditional inference, and propose a modification.

 We implicitly introduced the score test statistic $\bm g^T(\bm Y - \hat{\bm \mu})$, where $\hat{\bm \mu}$ is the maximum likelihood estimate of $\bm \mu$ under the null hypothesis, solved by $\bm U_{\bm \beta} = \bm 0$. Rather than approximating the distribution of this test statistic directly, the common procedure for score test statistics in the presence of nuisance parameters is to use conditional inference by conditioning $U_{\gamma}$ on $\bm U_{\bm \beta} = \bm 0$.
 
 Other methods for approximate conditional inference in the presence of nuisance parameters include orthogonal parametrization \citep{cox1987parameter} and projective methods \citep{waterman1996}. The first-order projective score, perhaps better known as the \textit{efficient score}, is for our model (Equation \eqref{model}) defined by 
\begin{align*}
\tilde{U}_{\gamma} &= U_{\gamma} - \bm F_{\gamma \bm \beta} F_{\bm \beta \bm \beta}^{-1} \bm U_{\bm \beta}.
\end{align*}
As noted by \cite{bickel1993efficient}, the efficient score may be interpreted in general as the score corresponding to a reparameterization $(\bm \beta, \gamma) \rightarrow (\bm \alpha, \gamma)$, by letting $\bm \beta(\bm \alpha,\gamma) = \bm \alpha -  F_{\bm \beta \bm \beta}^{-1}  \bm F_{\gamma \bm \beta}^T\gamma$. With this reparameterization of the logistic regression model, $\logit(\mu_i) = \bm x^{\text T}_i \bm \beta(\bm \alpha,\gamma)  + \gamma g_i = \bm x^\text T_i\bm\alpha+\gamma\tilde g_i$, where $\tilde g_i=g_i-\bm x_i^\text TF_{\bm \beta \bm \beta}^{-1} \bm F_{\gamma \bm \beta}^\text T$. Let $\tilde{F}$ denote the expected Fisher information of $\tilde{\bm U} = \smash{\begin{pmatrix}\tilde{\bm U}_{\bm \alpha}^\text T&\tilde{U}_\gamma\end{pmatrix}}^\text T$, the reparameterized score vector. With this reparameterization, the parameter $\gamma$ and the nuisance parameters $\bm \alpha$ are locally information orthogonal at $\gamma = 0$, which means that $\tilde{\bm F}_{\bm \alpha \gamma}$ and $\tilde{\bm F}_{\gamma \bm \alpha}$ in the expected Fisher information $\tilde{F}$ are zero-vectors (see e.g. \cite{lindsey1996parametric}). In this case, asymptotically $\tilde{\bm U}$ has a normal distribution, however additionally $\Cov(\tilde{\bm U}_{\bm \alpha}(\hat{\bm \mu}),\tilde{U}_{\gamma}(\hat{\bm \mu})) \rightarrow \bm 0$ when $\gamma = 0$ and $\bm \mu = \hat{\bm \mu}$. With $\tilde{\bm U}$ asymptotically multivariate normal, so will $\tilde{\bm U}_{\bm \alpha}$ and $\tilde{U}_{\gamma}$ (univariate) be. As covariance equal to zero for two normal distributed random variables implies independence, this means that the statistic of $\tilde{U}_{\gamma}$ conditional on $\tilde{\bm U}_{\bm \alpha} = \bm 0$ is asymptotically the same as the unconditional distribution of $\tilde{U}_{\gamma}$ when the null hypothesis is true with $\hat{\bm \mu}$ treated as a plug-in constant for $\bm \mu$.

In our case with expected Fisher information given in \eqref{fisher},
\begin{align*}
\begin{split}
&\tilde{U}_{\gamma} = \bm g^\text T(\bm Y - \bm\mu)-\bm g^\text TWX (X^\text TWX)^{-1}X^\text T(\bm Y-\bm \mu) \\
&= (\bm g^\text T-\bm g^\text TWX (X^\text TWX)^{-1}X^\text T) (\bm Y-\bm\mu) \\
&=(\bm g - X (X^\text TWX)^{-1}X^\text T W \bm g)^\text T (\bm Y-\bm\mu) \\
&= \tilde{\bm g}^\text T (\bm Y-\bm\mu),
\end{split}
\end{align*}
with $\tilde{\bm {g}} = \bm g - X(X^TW X)^{-1}X^T W \bm g$ the vector of all components $\tilde{g}_i$, and defined as in \cite{dey_fast_2017}. Observe that when $\bm U_{\bm \beta} = X^\text T(\bm Y-\bm \mu) = \bm 0$, the observed efficient score, $\tilde{u}$, is equal to $u$, the original observed score. Moreover, $E(\tilde{U}_{\gamma}|\tilde{\bm U}_{\bm \alpha} = \bm 0) = E(U_{\gamma}|\bm U_{\bm \beta} = \bm 0)= 0$, and $\Var(\tilde{U}_{\gamma}|\tilde{\bm U}_{\bm \alpha} = \bm 0) = \Var(U_{\gamma}|\bm U_{\bm \beta} = \bm 0) = \tilde{\bm g}^\text T W \tilde{\bm g}$ with $\tilde{\bm U}_{\bm \alpha} = \bm U_{\bm \beta}$ under the null hypothesis. At last, observe that asymptotically as $\hat{\bm \mu} \xrightarrow{p} \bm \mu$ under the null hypothesis,

\begin{align*}
\begin{split}
&\Cov(\tilde{U}_{\gamma}(\hat{\bm \mu}),\tilde{\bm U}_{\bm \alpha}(\hat{\bm \mu})) = F_{\gamma \bm \alpha}(\hat{\bm \mu}) =  E\left(\tilde{U}_{\gamma}(\hat{\bm \mu}) \tilde{\bm U}_{\bm \alpha}(\hat{\bm \mu})^\text T\right) \\
&= E\left(\left(\bm g - X \left(X^\text T\hat{W}X\right)^{-1}X^\text T \hat{W} \bm g\right)^\text T \left(\bm Y-\hat{\bm\mu}\right)\left(\bm Y-\hat{\bm \mu}\right)^\text T X\right) \\
&= \bm g^\text T E\left( \left(\bm Y-\hat{\bm\mu}\right)\left(\bm Y-\hat{\bm \mu}\right)^\text T \right)X - E\left(\bm g^\text T \hat{W} X \left(X^\text T\hat{W}X\right)^{-1} X^\text T\left(\bm Y-\hat{\bm\mu}\right)\left(\bm Y-\hat{\bm \mu}\right)^\text T \right)X \\
&\rightarrow \bm g^\text T WX -\bm g^\text T W X \left(X^\text TWX\right)^{-1} X^\text T WX = \bm 0^T.
\end{split}
\end{align*}
Hence, we have shown indeed that $\tilde{U}_{\gamma}$ and $\tilde{\bm U}_{\bm \alpha}$ are asymptotically independent under the null hypothesis.

Under the null hypothesis, using $\tilde{\bm U}$ leads asymptotically to the same \textit{unconditional} inference of $\tilde{U}_{\gamma}(\hat{\bm \mu})$ as when using the \textit{conditional} inference of $U_{\gamma}$ given $\bm U_{\bm \beta} = \bm 0$. In other words, $f(\tilde{U}_{\gamma}) \xrightarrow{d} N(0, \tilde{F}_{\gamma \gamma}(\hat{\bm \mu}))$, with $\tilde{F}_{\gamma \gamma}(\hat{\bm \mu})$ given in \eqref{condvar}. However, this will still be inaccurate for an imbalanced response and a skewed covariate of interest. Under this framework, we interpret the test proposed by \cite{dey_fast_2017} as a two-step approach, where the first step is to apply the efficient score, and in the second step the corresponding unconditional statistic is approximated by a single saddlepoint method via the univariate cumulant generating function of $\tilde{U}_{\gamma}$, given by
\begin{align*}
K(t) = \sum_{i=1}^n \ln(1 - \hat{\mu}_i + \hat{\mu}_i e^{\tilde{g}_i t}) - t \, \tilde{\bm g}^T \hat{\bm \mu}.
\end{align*}
Since such a two-step approach does not require a double saddlepoint approximation, this method is computationally more efficient. In \cite{dey_fast_2017}, the efficient score test statistic is assumed to have a continuous distribution. However, when $g_i \in (0,1,2)$, the efficient score test statistic in fact has a lattice distribution. Therefore, we propose to use a continuity correction. Similarly to the continuity-corrected double saddlepoint method outlined in the previous section, left-tail probabilities are estimated as in Equation \eqref{eq:BN}, now with 
\begin{align*}
w = \text{sgn}(\hat{t}) \sqrt{2(\hat{t} (u_{\gamma} - 1/2) - K(\hat{t}) )}, \text{ and } v = 2 \text{sinh}(\hat{t}/2) \sqrt{K''(\hat{t})},
\end{align*}
where $\hat{t}$ is the saddlepoint obtained by solving $K'(\hat{t}) = u_{\gamma} - 1/2$. Furthermore, we apply the same algorithm for obtaining two-sided $p$-values as in Section \ref{twosidedStrategy}.

\section{Comparison of methods}

For a specified significance level $\alpha$, a \textit{valid} test satisfies $P(\text{type I error}) \leq \alpha$. In our setting, we find it relevant to distinguish between conditional and overall (unconditional) validity. To clarify what is meant by this, consider a simple logistic regression model with no nuisance covariates (intercept only model). The covariate vector $\bm g$ is fixed while the response vector $\bm Y$ is random. Under the null, $Y_i \sim \text{binom}(1,\mu)$ for all $i = 1, \ldots, n$, where $\mu = \text{exp}(\beta_0)/(1+\text{exp}(\beta_0))$. For a particular realization $\bm y$, the observed score test statistic $u_{\gamma} = \bm g^T(\bm y - \hat{\bm \mu}) = \bm g^T(\bm y - \bar{y}\bm 1)$ may be compared to the conditional null distribution of $U_{\gamma}$, i.e. the distribution of $\bm g^T(\bm Y - \bar{y} \bm 1)$ given that $\bm Y$ is restricted by $\sum_i Y_i = n \bar{y} = v$ (Observation \ref{obs:exact1}). Thus, for all datasets in which the realization $\bm y$ satisfies $\sum_i y_i = v$, a test is \textit{conditionally} valid only when $P(\text{type I error}| \sum_i Y_i = v) \leq \alpha$. On the other hand, the \textit{overall} probability of type-1 error is given by 
\begin{align}
\sum_v \left[ P\left( \text{type I error} \mid \sum_{i=1}^n Y_i = v \right) P\left(\sum_{i=1}^n Y_i = v\right) \right].
\label{eq:t1error_interceptmodel}
\end{align}

A test that is conditionally valid for all $v$, will also be valid overall. The exact test derived in Observation \ref{obs:exact1} satisfies this property. An approximation to the exact test may be conditionally valid for some $v$, but invalid overall, or valid overall but conditionally invalid for some $v$. In the case with nuisance covariates, equation \eqref{eq:t1error_interceptmodel} may be generalized to:

\begin{align*}
 \sum_{X, \bm y \  : \  \bm U_{\bm \beta} = \bm 0} \left[P\left( \text{type I error} \mid \bm U_{\bm \beta} = \bm 0 \right) P\left( \bm U_{\bm \beta} = \bm 0\right)\right].   
\end{align*}

To evaluate the performance of our proposed methods, we consider both conditional and overall validity for models where the exact test is available. Approximation methods are evaluated based on their ability to control the overall type I error rate as well as the proportion of tests that are conditionally invalid.

\subsection{Intercept model}
In this section, we consider the intercept model with no nuisance parameters. We compare two discrete and two continuous conditional inference approximation methods with the exact test. The discrete methods are the double saddlepoint method with continuity correction as described in section 3, henceforth termed DSPA-CC, and the continuity-corrected single saddlepoint method based on the efficient score as described in section 4, henceforth termed ESPA-CC. The continuous methods are the normal approximation and the single saddlepoint method based on the efficient score (henceforth termed ESPA). To the best of our knowledge, the ESPA method mimics the SPA-test of \cite{dey_fast_2017} as implemented in the SPA-package in R. We present a simple example in order to highlight some of the key differences between the methods. 

\begin{figure}[t!]
\centering
\includegraphics[width=1\textwidth]{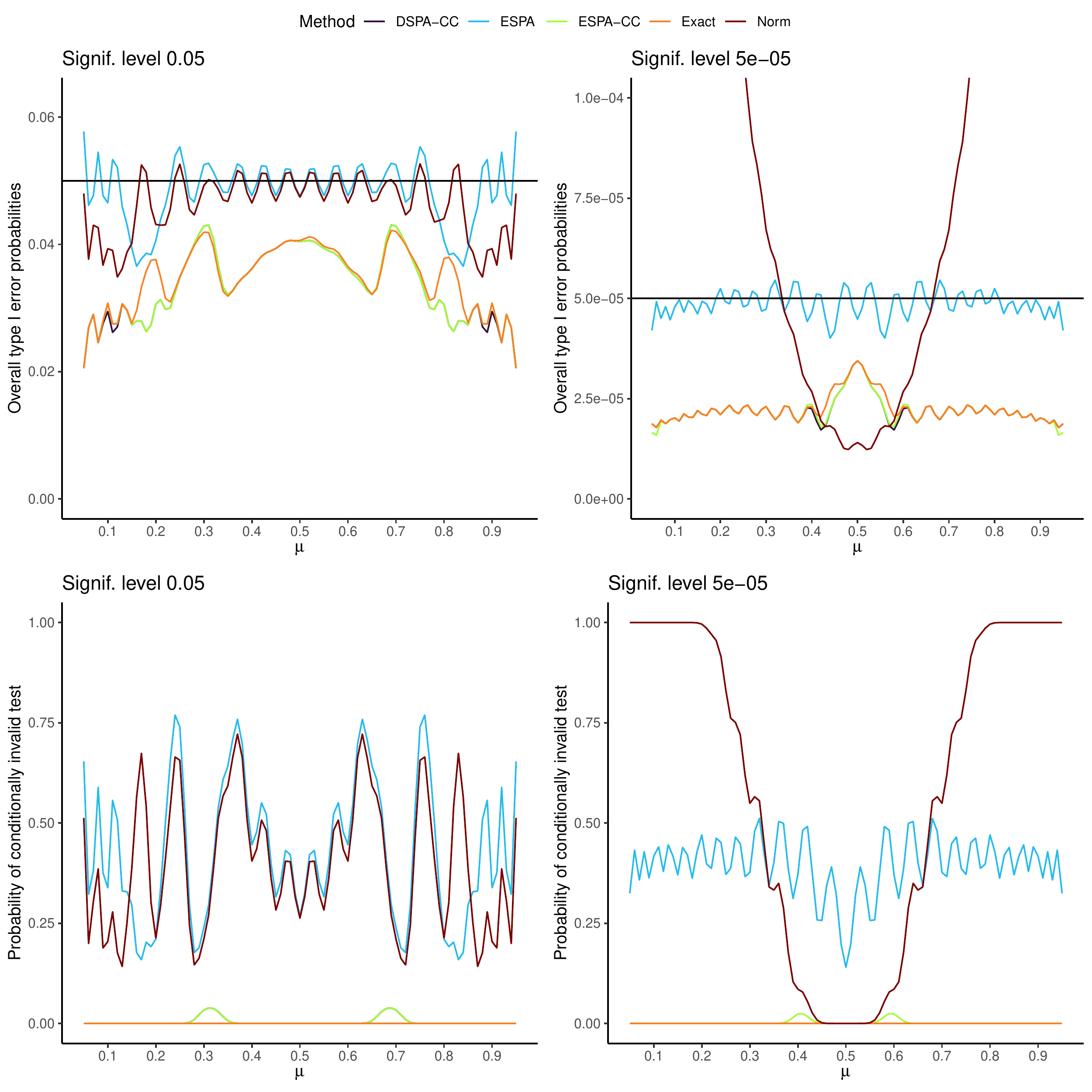}
\caption{Exact overall type I error probabilities as well as probability of conditionally invalid tests for the different approximations methods for the distribution of the score test statistic, for different values of $\mu$ in the intercept model. We compare with the exact test using the known distribution of the score test statistic.}
\label{fig:t1error_simple}
\end{figure}

Let $n = 1000$ and let $\bm g$ be the covariate vector with $n_0 = 980$ and $n_1 = 20$ and $n_2 = 0$. Without specifying what $\mu$ is, we first calculate the probabilities $P\left(\text{type I error} \mid \sum_{i=1}^n Y_i = v\right)$, for all $v = 1, 2, \ldots, n-1$. For a particular realization $v$, and discrete sample space within the support $[u_L, u_U]$ of the conditional null distribution of $U_{\gamma}$, where $u_L$ and $u_U$ need not be integers, we obtain the rejection region $\{u_L, \ldots, c_L\} \cup \{c_U, \ldots, u_U\}$ of the exact test. This can be achieved by a grid search from the left to obtain $c_L$ as well as a separate grid search from the right to obtain $c_U$ since the probability distribution is not symmetric. Then, $P\left( \text{type I error} \mid \sum_{i=1}^n Y_i = v \right) = P(U_{\gamma} \leq c_L \cup U_{\gamma} \geq c_U \mid \sum_{i=1}^n Y_i = v)$. For the approximation methods DSPA-CC, SPA-CC and SPA, we similarly use a grid search to identify lower ($c^*_L$) and upper ($c^*_U$) critical values that lead to rejection at the specified significance level. For the normal approximation, we obtain a critical value $c^*$ from the normal distribution with mean 0 and variance $\frac{v}{n}(1-\frac{v}{n}) \left[ n_0(0 - \frac{n_1}{n})^2 + n_1( 1 - \frac{n_1}{n})^2 \right]$, and then obtain the proper lower and upper critical values by the nearest grid points $c_L^*$ and $c_U^*$ to $-c^*$ and $c^*$ such that $c^*_L \leq -c^*$ and $c^*_U \geq c^*$. Then, for rejection regions  $\{u_L, \ldots, c_L^*\} \cup \{c_U^*, \ldots, u_U\}$, we calculate the exact conditional probability of erroneously rejecting the null hypothesis using the different approximation methods. For a specified value of $\mu$, we obtain probabilities $P\left(\sum_{i=1}^n Y_i = v\right)$ for each observed $v$. The overall probability of type I error can be computed according to Equation \eqref{eq:t1error_interceptmodel}. In addition, the probability of a conditionally invalid test for each method and for each $\mu$ can be computed by observing which values $v$ where $P(\text{type I error}\mid \sum_i Y_i = v) > \alpha$, and add together the probabilities $P\left(\sum_{i=1}^n Y_i = v\right)$ for each such $v$. See Figure \ref{fig:t1error_simple}. 

From this example, we make four observations;
\begin{enumerate}
 \item The exact test is always conservative (see Figure \ref{fig:t1error_simple}). When a significance level $\alpha$ is specified, the discrete nature of the test results in an achieved significance level less than $\alpha$. This observation is of course a well-known issue with discrete test statistics. 
 \item Both of the discrete approximations (DSPA-CC and SPA-CC) closely resemble the exact test in terms of overall type-1 error rates (Figure \ref{fig:t1error_simple}). At significance level $\alpha = 0.05$, both methods gave conditionally invalid tests in four situations; $v=301$, $v=325$, $v=675$, and $v=699$. For instance for $\mu = 0.31$ and $\mu = 0.69$, this results in probabilities $ \approx 0.04$ of sampling a dataset where these methods are conditionally invalid. At significance level $\alpha = 5\cdot 10^{-5}$, DSPA-CC is conditionally valid for any $v$, while SPA-CC is conditionally invalid for $v = 406 $ and $v = 594$. For instance for $\mu = 0.41$ and $\mu = 0.59$, this results in a slight probability ($\approx 0.02$) of sampling a dataset where the PSA-CC method is conditionally invalid.
 \item Even at significance level $\alpha = 0.05$, the normal approximation is invalid for different $\mu$-values (Figure \ref{fig:t1error_simple}). For significance level $5 \cdot 10^{-5}$, the normal approximation is valid when the response is balanced ($\mu \approx 0.5$). However, for skewed responses (small or large $\mu$), the normal approximation becomes severely unreliable. At significance level $\alpha = 0.05$, the normal approximation was conditionally invalid in around $40\%$ of possible realizations of $\sum_i Y_i$. At significance level $\alpha = 5\cdot 10^{-5}$, this number had increased to around $64\%$. The majority of situations where the normal approximation was conditionally invalid was for small or large number of cases $v$, which is in-line with the observations made of overall type-1 error rates for skewed responses (Figure \ref{fig:t1error_simple}). 
 
 \item The SPA method is less conservative than the exact test, and at times anti-conservative. At significance level $\alpha = 0.05$, the SPA method was conditionally invalid around $43\%$ of possible realizations of $\sum_i Y_i$, and at significance level $\alpha = 5 \cdot 10^{-5}$, the SPA method was conditionally invalid around $39\%$ of situations. As opposed to the normal approximation method where invalid tests clustered towards skewed response distributions, the SPA method fluctuates relatively evenly between conditionally valid and conditionally invalid as the number of cases $v$ increases for both significance levels 0.05 and $5 \cdot 10^{-5}$. Therefore, the test is approximately equally good at any $\mu$ (Figure \ref{fig:t1error_simple}). Furthermore, the absolute differences in type-1 error rate control improves as the significance level decreases. This observation has a simple explanation. For some data sets, the SPA method yields the same critical region as the exact test, while at times the critical region is shifted by as little as one unit ($c^*_U = c_U - 1$ or $c^*_L = c_L + 1$). At a significance level of $\alpha = 0.05$, this shift can result in a substantial inflation in type-1 error rates, while at small significance levels, point probabilities are of such small magnitudes that the shift is less notable. As critical regions oscillate between correct and slightly shifted, conditional type-1 error rates oscillate above and below $\alpha$, and averaging out to produce an overall type-1 error rate $\approx \alpha$.
\end{enumerate}

\subsection{Simulations of genetic association studies with an imbalanced response}

The purpose of the following simulation study is to compare methods in a setting resembling a genome-wide association study with an imbalanced response, in which exact tests are not available. The simulation set-up is motivated by \cite{dey_fast_2017} by conditioning on the number of cases, and we estimate the type I error rate \textit{conditional} on the number of cases. The sample size considered is $n = 20.000$, with case proportion $2\%$ and $0.2\%$. We consider the logistic regression model
\begin{align*}
\text{logit}(\mu_i) = \beta_0 + x_{i,1} + x_{i,2} + \gamma g_i,
\end{align*}
with $X_{1}\sim \text{Bernoulli}(0.5)$, $X_{2} \sim N(0,1)$ and $G \sim \text{binom}(2,\text{MAF})$ with the MAF taking the values $0.05$, $0.005$, $0.0005$ and $0.00025$. Since we are evaluating validity of tests, we set $\gamma = 0$. Finally, we set $\beta_0 = -5.6$ such that the disease prevalence is $1 \%$ in the population. 

\begin{figure}[h]
\centering
\includegraphics[width=0.8\textwidth]{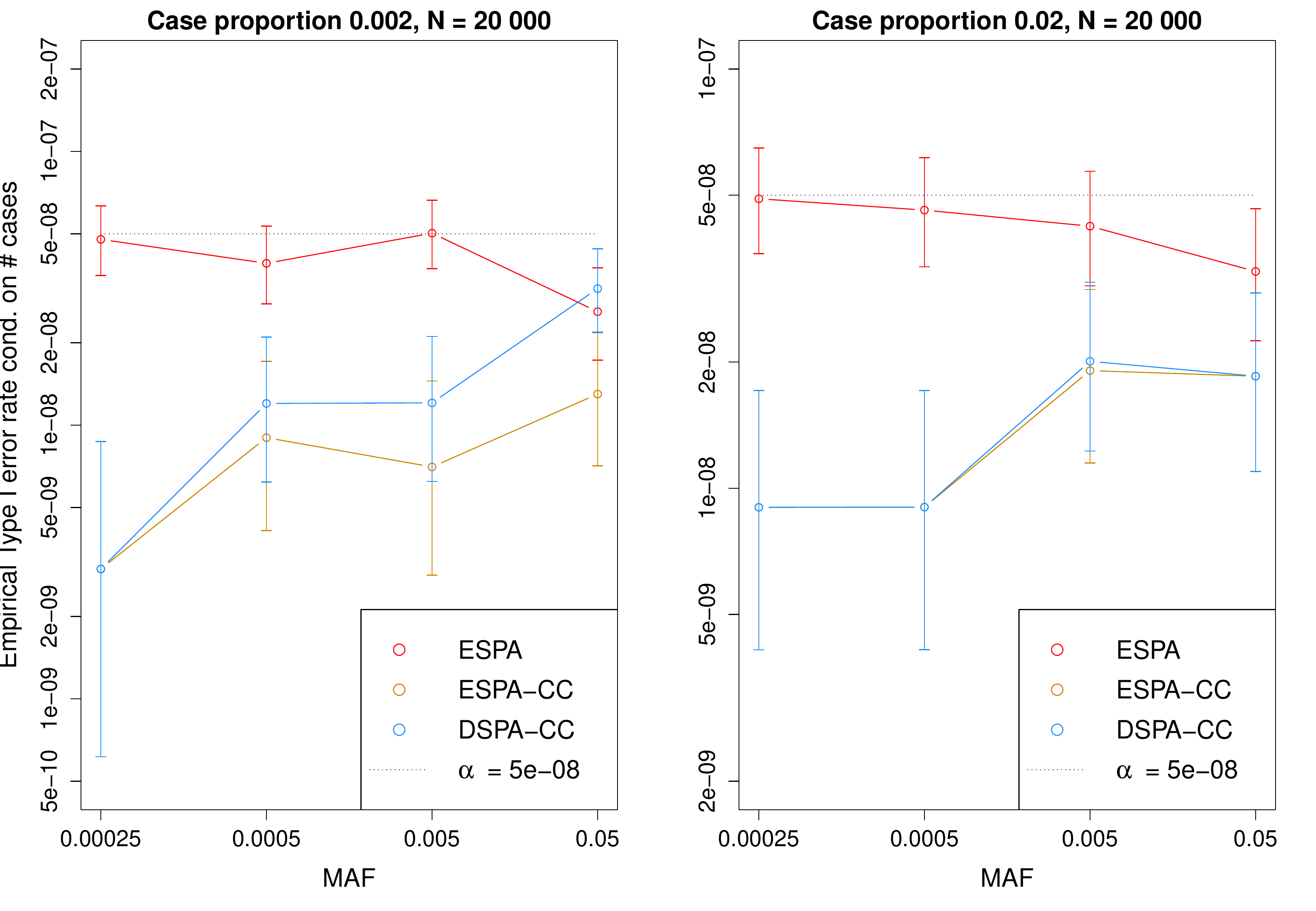}
\caption{Approximated expected type I error rates - conditional on the number of cases - for ESPA, ESPA-CC and DSPA-CC from simulations with case proportions 0.02 and 0.002, and for small MAFs when nuisance covariates are included.}
\label{TIEUB}
\end{figure}

The covariates $x_{i,1}$ and $x_{i,2}$ are sampled conditionally  on their respective phenotype value $y_i$, while the genotype value is sampled independently of this under the null hypothesis. See Supplementary File for details. This ensures that the number of cases is equal for all simulations. For each set of case proportion and MAF, we simulate $10^9$ data sets and record the amount of times the null hypothesis is rejected at the $\alpha = 5\cdot 10^{-8}$ significance level when using (1) the double saddlepoint approximation with continuity correction (DSPA-CC), (2) a continuity-corrected univariate saddlepoint approximation based on the efficient score (ESPA-CC), and (3) a continuous univariate saddlepoint approximation of the efficient score (ESPA). The resulting empirical type I error rates are presented in Figure \ref{TIEUB}, along with 95\% Clopper-Pearson confidence intervals. 

The simulation results closely follow the observations made in the previous section. The DSPA-CC and ESPA-CC are conservative (overall probability of type I error $< \alpha$), while the type I error rate of the ESPA method is $\approx \alpha$. The results are comparable with the pattern of conditionally invalid tests in Figure \ref{fig:t1error_simple}, specifically for the small case proportion, in that we sense a large fluctuation in the probability of invalid tests for ESPA, while both ESPA-CC and DSPA-CC have a small probability of invalid test, which is decreasing for decreasing MAF. We also observe a trend that the type I error rate, conditional on the number of cases, for EPSA is increasing for decreasing MAF. The simulation study with case proportion 0.002 serves to illustrate deviations between the DSPA-CC and ESPA-CC method, and we observe that the ESPA-CC is somewhat more conservative in this setting.

\section{Application to UK biobank data}

We consider a recent GWAS in the UK Biobank with motivation from \cite{rogne_gwas_2021}. The phenotype of interest is skin and soft tissue infections (SSTIs), and individuals are defined as cases if they have been hospitalized with main ICD-10 codes A46 (erysipelas), L03 (cellulitis and acute lymphangitis), or M72.6 (necrotizing fasciitis) in the period between the end of the recruitment period (2010-10-01) and April 2017 (2017-03-31). Individuals who had reported ICD-10 codes, or corresponding ICD-9 codes (035 and 729.4), before 2010-10-01 are removed as well as individuals with date of death reported after 2010-10-01 in the death register (see Data-Field 40000 in the UK Biobank data). As nuisance covariates we include age when attended assessment centre, genetic sex, and four principal components. To avoid complexities due to cryptic relatedness we only include unrelated individuals reported as Caucasians (achieved through Data-Field 22006 and 22020 in UK Biobank). The principal components are calculated using EIGENSOFT (version 6.1.4) SmartPCA \citep{price_principal_2006, patterson_population_2006}. Only directly genotyped SNPs are considered, and phenotype-independent quality control of the genetic data is completed using PLINK1.9, with details given in the Supplementary File. This results in a total of 293 964 individuals and 529 024 SNPs with 2051 individuals defined as cases and 291 913 controls, resulting in a case proportion of 0.7 \%.
All SNPs are first investigated by computing $p$-values using the normal approximation to the score test statistic. As this test is proven to be too optimistic, SNPs with $p$-values less than $\alpha = 5 \cdot 10^{-5}$ are investigated more thoroughly by computing  $p$-values using the DSPA-CC and ESPA-CC methods as implemented by us, as well as the SPA-test of \cite{dey_fast_2017}, denoted ESPA. \cite{dey_fast_2017} also propose a computationally more efficient approximation to their SPA-test by essentially assuming that the nuisance covariates are balanced. In a double saddlepoint setting, this assumption may be generalized to argue that the score vector $\bm U_{\bm \beta}$ approximately has a multivariate normal distribution under the null hypothesis. Taking a similar approach to \cite{dey_fast_2017}, we may partition the joint CDF of $\bm U_{\bm \beta}$ and $U_{\gamma}$ into a sum over all individuals with genotype value $g_i > 0$ and those with $g_i = 0$. For the latter sub-sample, the CGF simplifies to a CGF of the score vector $\bm U_{\bm \beta}^*$ including individuals with $g_i = 0$. Assuming that also $\bm U_{\bm \beta}^*$ is normal, this part of the joint CGF may be replaced by a normal CGF, and by pre-computing the variance of $\bm U_{\bm \beta}^*$, an approximated double saddlepoint method may be computed based only on the sub-sample individuals with genotypes $g_i > 0$. Details may be found in the Supplementary File. For comparative purposes, we also compute $p$-values based on the fastSPA method of \cite{dey_fast_2017} and our similar fastDSPA-CC approach.

Test results for the SNPs with the smallest normal-approximated $p$-values are given in Table \ref{GWAS_skin}. In this setting, we no longer know whether the null hypothesis is true or not for each variant. However, we expect only a tiny proportion of all variants where the null hypothesis is false. Even though no SNPs reached the significance level $\alpha = 5 \cdot 10^{-8}$, we see a pattern similar to the results for the intercept model and our simulation results. The normal approximation is the most optimistic,  followed by ESPA and fastSPA tests. The DSPA-CC test is more conservative, while the most conservative test is ESPA-CC. The fastDSPA-CC is slightly less conservative than DSPA-CC. The greatest difference between test results is observed for the SNP with a small minor allele frequency (rs113113104, MAF = 0.03). The difference between the $p$-values reduces for increasing MAFs. For the SNP rs566530 with MAF = 0.48, the SPA test gives a smaller $p$-value than the normal approximation, while the other methods give consistently larger $p$-values. 

\begin{table*}[ht]
\centering
    \caption{Results for GWAS of skin and soft tissue infections for common variants.}
    \scalebox{0.87}{
    \begin{tabular}{|l | l | l | l | l | l | l | l | l |}
    \hline
     \textbf{SNP} & \textbf{CHR} & \textbf{MAF} & \textbf{Norm. apx.} & \textbf{ESPA} & \textbf{fastSPA} & \textbf{SPA-CC} & \textbf{DSPA-CC} & \textbf{fastDSPA-CC}   \\ \hline
     rs113113104 & 6 & 0.03 & 2.39e-07 & 5.97e-07 & 6.04e-07 & 7.27e-07 & 7.10e-07 & 6.52e-07  \\ \hline
     rs6551253 & 3 & 0.28 & 8.38e-06 &  8.47e-06 & 8.78e-06 & 9.18e-06 & 9.00e-06 & 8.92e-06    \\
     \hline
     rs78404737 & 2 & 0.10 & 8.50e-06 & 9.63e-06 & 9.78e-06 & 1.08e-05 & 1.06e-05 & 1.00e-05  \\
     \hline
     rs78696065 & 7 & 0.02 & 8.80e-06 & 1.54e-05 & 1.55e-05 & 1.89e-05 & 1.87e-05 & 1.75e-05 \\
     \hline
     rs479947 & 6 & 0.11 &  1.19e-05 & 1.29e-05 & 1.33e-05 & 1.44e-05 & 1.42e-05 & 1.35e-05 \\
     \hline
     rs566530 & 6 & 0.48 & 1.46e-05 & 1.40e-05 & 1.48e-05 & 1.50e-05 & 1.47e-05 & 1.48e-05  \\
     \hline
     rs56355912 & 10 & 0.03 & 1.51e-05 & 2.16e-05 & 2.16e-05 & 2.57e-05 & 2.54e-05 & 2.38e-05 \\
     \hline 
     rs72733294 & 5 & 0.36 & 1.58e-05 & 1.60e-05 & 1.60e-05 & 1.72e-05 & 1.69e-05 & 1.69e-05 \\
     \hline
     rs11074743 & 16 & 0.40 & 1.69e-05 & 1.68e-05 & 1.71e-05 & 1.80e-05 & 1.77e-05 & 1.77e-05 \\
     \hline
     rs1562963 & 11 & 0.07 & 2.02e-05 & 1.99e-05 & 2.33e-05 & 2.26e-05 & 2.23e-05 & 2.13e-05 \\
     \hline
    \end{tabular}
    }
    \label{GWAS_skin}
\end{table*}

\subsection{Rare variants}

The difference between the methods becomes even larger when investigating rare variants. We consider the UK Biobank exome sequence data consisting of 45 596 unrelated individuals of European origin. We limit ourselves to White-British Caucasians using the same requirements for the definition of SSTIs as for the common variants. This results in a total number of 30 210 individuals to investigate with 210 individuals defined as cases, once again leading to a case proportion of about 0.7 \%. See Supplementary File for further information about quality control. The principal components are computed as for the common variants analysis, however separately on these 30 210 individuals. We will in addition only consider chromosome 6 as well as rare variants with a minimum minor allele count (MAC) equal to 3. The results are given in Table \ref{exome_skin}.

\begin{table*}[ht]
\centering
    \caption{Results for GWAS of skin and soft tissue infections for rare variants.}
    \scalebox{0.83}{
    \begin{tabular}{| l | l | l | l | l | l | l | l | l |}
    \hline
      \textbf{SNP} & \textbf{CHR} & \textbf{MAC} & \textbf{Norm. apx.} & \textbf{ESPA} & \textbf{fastSPA} &  \textbf{ESPA-CC} & \textbf{DSPA-CC} & \textbf{fastDSPA-CC}   \\ \hline
     6:26045407:G:A & 6 & 4  & 2.07e-36 & 4.31e-05 & 4.31e-05 & 2.2e-04 & 2.2e-04 &  2.2e-04  \\ \hline
     6:41097421:T:C & 6 &  4 & 2.21e-32 &  4.92e-05 & 4.92e-05 & 2.6e-04 & 2.6e-04 & 2.5e-04   \\
     \hline
     6:24852645:G:T & 6 & 4 & 1.37e-25 & 8.93e-05 & 8.93e-05 & 4.4e-04 & 4.3e-04 & 4.2e-04 \\
     \hline
     6:31772925:C:A & 6 & 5 & 6.36e-23 & 1.3e-04 & 1.3e-04 & 6.0e-04 &  6.0e-04 & 5.8e-04 \\
     \hline
     6:20402579:C:T & 6 & 3 & 4.19e-22 &  0.0020 & 0.0020 & 0.010 & 0.010 & 0.010 \\
     \hline
      6:132588925:C:T & 6 & 6 & 8.78e-22 & 1.5e-04 &  1.5e-04 &  6.9e-04 & 6.9e-04 & 6.7e-04 \\
     \hline
      6:17675831:G:A & 6 & 3 & 8.94e-22 & 0.0020 & 0.0020 & 0.010 &  0.010 & 0.010 \\
     \hline
      6:110960684:T:G & 6 & 3 & 2.05e-21 & 0.0017 & 0.0017 &  0.0049 & 0.0049 &  0.0049 \\
     \hline
    6:7894854:T:C & 6 & 16 &  1.88e-20 & 3.07e-05 & 3.073e-05 & 1.2e-04 &  1.2e-04 & 1.0e-04 \\
     \hline
     6:148514044:G:T & 6 & 3 & 1.94e-20 & 0.0022 & 0.0022 & 0.011 & 0.011 &  0.011 \\
     \hline
    \end{tabular}
    }
    \label{exome_skin}
\end{table*}

It is clear that the normal approximation to the score test statistic is very inaccurate in this setting. However, we also see that the difference between ESPA and the other saddlepoint approximations with continuity correction differ in about one order of magnitude. As a result, we expect the importance of the continuity correction to be most consequential for rare variants. Another observation is that ESPA-CC and DSPA-CC are practically identical in this case. We also see that the speed-up approximation methods are more accurate which can be explained by observing that the accuracy of the multivariate normal approximation of $\bf{U}_{\bf{\bm\beta}}^{*}$ in fastDSPA-CC, depends on the number of individuals with $g_i = 0$, which increases for decreasing MACs. The same applies for the approximation of the corresponding normal distribution in fastSPA.

\section{Discussion}

We have investigated different saddlepoint approximations for GWAS with binary phenotypes in order to achieve valid $p$-values. We have shown how the saddlepoint approximation introduced in \cite{dey_fast_2017} can be interpreted as a two-stage procedure in which one first applies the efficient score to approximate the conditional score test statistic as an unconditional statistic, and then perform single-saddlepoint approximation. We further show how to apply the double saddlepoint approximation to directly approximate the conditional score test statistic.

We distinguish between conditional and overall type I error rate. Taking into account both these measures, we conclude that continuity-corrected saddlepoint approximations are most appropriate in this setting. The continuity-corrected double saddlepoint approximation, DSPA-CC, and single-saddlepoint approximation, ESPA-CC, using the efficient score are both considered to perform well, however there are situations in which ESPA-CC is somewhat more conservative than DSPA-CC, indicating DSPA-CC to be somewhat more powerful.

There are additional continuity correction variants, and the one used here is called the \textit{second continuity correction}. A first and a third continuity correction are alternatives \citep{butler_2007}, and specifically the first continuity correction was also investigated with very similar results as when using the second continuity correction, however slightly more inaccurate when considering the intercept model, see Supplementary File.
An alternative saddlepoint approximation to the CDF of a random variable is the one introduced in \cite{lugannani_rice_1980}. This approximation gives the same results as the approximation by \cite{barndorff-nielsen_approximate_1990} in most situations. However, we observed under the simulations that when the case proportion and MAF approaches zero, the approximation by Lugananni and Rice is inaccurate, see Supplementary File. See for instance \cite{LRfails} for similar observations in a different application.

Consider the case where one wants to include imputed SNPs. For most imputation methods, the output for each imputed SNP is a probability that the minor allele count is equal to 0, 1 or 2, denoted $p_0, p_1$ and $p_2$. Then one must be aware of the fact that when the imputed genotype is set to be the expected minor allele count, $p_1 + 2p_2$, the score test statistic will no longer have a lattice distribution, and so continuity correction does no longer apply. However, to account for imputed SNPs in our method one can instead set the imputed minor allele count to be equal to the most likely allele count according to the imputation method.

Single-variant tests on rare variants are often low-powered, and therefore several region-based tests including several SNPs in the same genetic region have been proposed to gain power. However, many of these methods again rely on single-variant tests as building blocks, among them SKAT and ACAT \citep{skat,liu_acat_2019}. It is therefore essential that the single-variant tests are sufficiently accurate. Future work could be how the insight of the score test statistic introduced in this work would impact region-based tests.  

\section{Acknowledgements}
This research was supported by the Norwegian Research Council grant 272402 (PhD Scholarships at SINTEF) as well the funding for research stays abroad for doctoral and postdoctoral fellows financed by the Norwegian Research Council. The research has been conducted using the UK Biobank Resource under Application Number 32285. We thank the Yale Center for Research Computing for guidance and use of the research computing infrastructure. We thank the The Gemini Center for Sepsis Research for establishing cooperation with Yale School of Public Health.

\section{Code availability}
Source code is available at \url{https://github.com/palVJ/SaddlePointApproxInBinaryGWAS}.

\appendix

\section{Proofs of Observations \ref{obs:discrete_bounded}--\ref{obs:exact2}}\label{obsproofs}

\begin{proof}[Proof of Observation \ref{obs:discrete_bounded}]
When $g_i \in (0,1,2)$, we note that $\bm g^\text T \bm Y$ is an integer and $\bm g^\text T \bm\mu$ a constant, so that $U_{\gamma}=\bm g^\text T \bm Y-\bm g^\text T \bm\mu$ has support on a subset of a lattice with step~1. The minimum is obtained for $\bm Y=\bm0$ and the maximum for $\bm Y=\bm1$ (a vector of ones), and the result follows.
\end{proof}

\begin{proof}[Proof of Observation \ref{obs:exact1}]
We assume throughout the proof that the null hypothesis is true, $\gamma=0$. Denote by $V_j$ the sum of responses $Y_i$ among individuals with genotype $g_i = j$, \ $j=0$, 1, 2, and let $V = V_0 + V_1 + V_2 = \sum_{i=1}^n Y_i$ be the total sum of responses. With this notation, $U_{\gamma}= V_1+\nobreak  2V_2 -\nobreak (n_1 +\nobreak 2n_2) \mu$, and $U_\beta=V-n\mu$, so that the condition $U_{\beta} = 0$ is equivalent to $V = n \mu$.

The $V_j$ are independent, and $V_j$ is binomially distributed with parameters $n_j$ and~$\mu$, \ $j = 0$, 1, 2, and~$V$ is binomially distributed with parameters $n$ and $\mu$. Assume that $v_0+v_1+v_2=n\mu$ with $v_j$ in the support of $V_j$. Then
\begin{multline*}
 P(V_0=v_0,V_1=v_1,V_2=v_2\mid V = n\mu\bigr)
  = \frac{P(V_0 = v_0)P(V_1 = v_1) P(V_2 = v_2)}{P(V =  n\mu)}  \\
 = \frac{ \binom{n_0}{v_0} \mu^{v_0}(1-\mu)^{n_0 - v_0}  \binom{n_1}{v_1} \mu^{v_1}(1-\mu)^{n_1 - v_1}  \binom{n_2}{v_2} \mu^{v_2}(1-\mu)^{n_2 - v_2} }{\binom{n}{n\mu} \mu^{n\mu}(1-\mu)^{n -  n\mu}} 
 = \frac{ \binom{n_0}{v_0}  \binom{n_1}{v_1} \binom{n_2}{v_2}  }{\binom{n}{ n\mu} },
\end{multline*}
a trivariate hypergeometric probability.

Now, $P(U_{\gamma} = u \mid U_{\beta} = 0) =P\bigl(V_1 + 2V_2 =u^*\bigm| V=n\mu\bigr)$ can be found by summing the above probabilities over $(v_0,v_1,v_2)\in S$. This gives the first sum of the Observation. The more explicit second version of the sum is obtained by solving the two equations in the definition of $S$ for $v_0$ and $v_1$ in terms of $k=v_2$. The limits of the sum is determined by the inequalities in the definition of $S$.
\end{proof}

\begin{proof}[Proof of Observation \ref{obs:exact2}]
We assume throughout the proof that the null hypothesis is true, $\gamma=0$. Denote by $V_j$ the sum of responses $Y_i$ among individuals with $x_i = 0$ and genotype $g_i = j$, \ $j=0$, 1, 2, and let $V = V_0 + V_1 + V_2$. Define similar sums $W_j$ and $W$ for individuals with $x_i = 1$. With this notation, $U_\gamma=V_1+2V_2-(l_1+2l_2)\mu_0+W_1+2W_2-(m_1+2m_2)\mu_1$, and $\bm U_{\bm \beta}^\text T=\begin{pmatrix}V+W-l\mu_0-m\mu_1&W-m\mu_1\end{pmatrix}$, so that the condition $\bm U_{\bm \beta}=\bm0$ is equivalent to $V=l\mu_0$ and $W=m\mu_1$.

All the $V_j$ and $W_j$ are independent, and $V_j$ is binomially distributed with parameters $l_j$ and $\mu_0$, and~$W_j$  with parameters $m_j$ and $\mu_1$, \ $j=0$, 1, 2. As in the proof of Observation~\ref{obs:exact1}, the conditional point probabilites of $(V_0,V_1,V_2)$ given $V=l\mu_0$ and $(W_0,W_1,W_2)$ given $W=m\mu_1$ are trivariate hypergeometric probabilities, and by independence of the two triples, the conditional joint point probability is the product of the two. Then $P(U_{\gamma} = u \mid\bm U_{\bm \beta} = \bm 0)$
can be found by summing those probabilities over $\bm s\in S$.
\end{proof}

\section{Support of  the conditional score test statistic}\label{supportCondScore}
Consider the score test statistic of $U_{\gamma}$ conditional on $\bm U_{\bm \beta} = \bm 0$, given by $\bm g^T(\bm Y - \bm \hat{\bm \mu})$.
We have $-\hat{\bm\mu}\leq\bm Y-\hat{\bm\mu}\leq\bm1-\hat{\bm\mu}$ (elementwise inequalities), where $\bm1$ is a vector of ones. Since all $g_i \ge 0$, premultiplying the inequalities with $\bm g^\text T$ gives bounds on the support of $\bm g^T(\bm Y - \bm \hat{\bm \mu})$:
\begin{equation}
 -\bm g^\text T\hat{\bm\mu}\leq U_{\gamma} \leq\bm g^\text T(\bm1-\hat{\bm\mu}).\label{ubounds}
\end{equation}
The first equality holds when $\bm g^\text T\bm Y=0$ and the second when $\bm g^\text T\bm Y=\bm g^\text T\bm 1$. However, this combination is not achievable if it does not satisfy $\bm U_{\bm \beta} = \bm X^T (\bm Y - \hat{\bm \mu}) = \bm 0$. Specifically, the minimal and maximal achievable values of the conditional score test statistic is given by the constraint optimization problems:

\begin{align*}
    \min(U_{\gamma}) = \min_{\bm y} \quad & \bm g^T(\bm y - \bm \hat{\bm \mu}) \\ 
    \text{such that} \quad &  X^T (\bm y - \hat{\bm \mu}) = \bm 0,
\end{align*}
and
\begin{align*}
    \max(U_{\gamma}) = \max_{\bm y} \quad & \bm g^T(\bm y - \bm \hat{\bm \mu}) \\ 
    \text{such that} \quad &  X^T (\bm y - \hat{\bm \mu}) = \bm 0.
\end{align*}

As an example, consider the intercept model with $n=1000$ and $\bm g$ as in Section 5.1 with $n_0= 980$, $n_1 = 20$ and $n_2 = 0$ as well as the observation $\sum_{i=1}^{1000} Y_i = 10$. Then $\hat{\mu}_i = 10/1000 = 0.01$ satisfies $U_{\beta_0} = \sum_{i=1}^{1000} (Y_i-\mu_i) = 0$. Then the minimum achievable value is indeed $\min(U_{\gamma}) = -\bm g^T \hat{\bm \mu} = -0.2$, since we may have a combination where $Y_i = 0$ for all $g_i > 0$, and still get $\sum_{i=1}^{1000} Y_i = 10$. However, $\max(U_{\gamma}) = 10 -\bm g^T \hat{\bm \mu} = 9.8$ since $\bm g^T \bm Y$ can be no larger than the combinations where $g_i = 1$ for all $Y_i = 1$, which can only occur ten times in order to satisfy $\sum_{i=1}^{1000} Y_i = 10$.

\newsavebox{\boks} 
\savebox{\boks}{\textbf{\Large Solution to $\nabla_{\bm t_{\bm \beta}}K_{\bm \beta}(\tilde{\bm t}_{\bm \beta}) = \bm 0$}} 
\section{\texorpdfstring{\usebox{\boks}}{K-gradient}}\label{Sol_t_tilde}

Given the marginal cumulant generating function of $\bm U_{\bm \beta}$, defined by $K_{\bm \beta}(\bm t_{\bm \beta})$ (a function of $d$ variables) with

\begin{equation}
K_{\bm \beta}(\bm t_{\bm \beta}) = \sum_{i=1}^n \ln(1-\mu_i+\mu_i \exp(\mathbf{x}^{T}_{i} \bm t_{\bm \beta}))-\bm t_{\bm \beta}^T X^T \boldsymbol{\mu},
\label{Kbeta}    
\end{equation}
and corresponding gradient
\begin{equation}
\nabla_{\bm t_{\bm \beta}} K_{\bm \beta}(\bm t_{\bm \beta}) = \sum_{i=1}^n \mu_i \mathbf{x}_i\left(\frac{1}{(1-\mu_i)\exp(-\mathbf{x}^{T}_{i}\bm t_{\bm \beta})+\mu_i}-1\right).
\label{kprime}
\end{equation}

First, one can easily observe that $\tilde{\bm t}_{\bm \beta} = \bm 0$ is a solution to $\nabla_{\bm t_{\bm \beta}}K_{\bm \beta}(\bm t_{\bm \beta}) = \bm 0$. Second, if one can prove that the CGF is a convex function, then $\tilde{\bm t}_{\bm \beta} = \bm 0$ is a unique solution to $\nabla_{\bm t_{\bm \beta}}K_{\bm \beta}(\bm t_{\bm \beta}) = \bm 0$.\\
\begin{proof}
 In fact, convexity of a cumulant generating function with \textit{any} random variable $\bm U$, $K(\bm t)=\ln E(e^{\bm t^T \bm U})$, in general follows from the Hölder inequality, $E(\lvert X\rvert^c\lvert Y\rvert^{1-c})\leq(E\lvert X\rvert)^c(E\lvert Y\rvert)^{1-c}$ for all $c$ in $(0,1)$, where $X$ and $Y$ are random variables. A function  $f$ is convex if $f(c\bm t_1+(1-c)\bm t_2)\leq cf(\bm t_1)+(1-c)f(\bm t_2)$ for all $c$ in $(0,1)$. Now,
\[\begin{split}
 K(c\bm t_1+(1-c)\bm t_2)&=\ln Ee^{(c\bm t_1+(1-c)\bm t_2)^\text T\bm U}
  =\ln E\big(e^{c\bm t_1^\text T\bm U}e^{(1-c)\bm t_2^\text T\bm U}\big)\\
 &\leq\ln\bigl(\bigl(Ee^{\bm t_1^\text T\bm U}\bigr)^c\bigl(Ee^{\bm t_2^\text T\bm U}\bigr)^{1-c}\bigr)
  =c\ln Ee^{\bm t_1^\text T\bm U}+(1-c)\ln Ee^{\bm t_2^\text T\bm U}\\
  &=cK(\bm t_1)+(1-c)K(\bm t_2),
\end{split}\]
showing that $K$ is convex.
\end{proof}

\printbibliography

@article{dey_fast_2017,
	title = {A Fast and Accurate Algorithm to Test for Binary Phenotypes and Its Application to {PheWAS}},
	volume = {101},
	%url = {https://www.ncbi.nlm.nih.gov/pmc/articles/PMC5501775/},
	number = {1},
	journal = {American Journal of Human Genetics},
	author = {Dey, Rounak and Schmidt, Ellen M. and Abecasis, Goncalo R. and Lee, Seunggeun},
	month = jul,
	year = {2017},
	pages = {37--49},
}

@article{ma_recommended_2013,
	title = {Recommended joint and meta-analysis strategies for case-control association testing of single low-count variants},
	volume = {37},
	number = {6},
	journal = {Genetic epidemiology},
	author = {Ma, Clement and Blackwell, Tom and Boehnke, Michael and Scott, Laura J.},
	month = sep,
	year = {2013},
	pages = {539--550},
}

@book{butler_2007, 
    series={Cambridge Series in Statistical and Probabilistic Mathematics}, title={Saddlepoint Approximations with Applications}, %DOI={10.1017/CBO9780511619083},
    publisher={Cambridge University Press}, author={Butler, Ronald W.}, 
    year={2007}, 
    collection={Cambridge Series in Statistical and Probabilistic Mathematics},
}

@article{firth1993bias,
  title={Bias reduction of maximum likelihood estimates},
  author={Firth, David},
  journal={Biometrika},
  volume={80},
  number={1},
  pages={27--38},
  year={1993},
  publisher={Oxford University Press}
}

@article{rogne_gwas_2021,
	title = {{GWAS} {Identifies} {LINC01184}/{SLC12A2} as a {Risk} {Locus} for {Skin} and {Soft} {Tissue} {Infections}.},
	journal = {J Invest Dermatol},
	author = {Rogne, T. and Liyanarachi, K. V. and Rasheed, H. and Thomas, L. F. and Flatby, H. M. and Stenvik, J. and Løset, M. and Gill, D. and Burgess, S. and Willer, C. J. and Hveem, K. and Åsvold, B. O. and Brumpton, B. M. and DeWan, A. T. and Solligård, E. and Damås, J. K.},
	month = mar,
	year = {2021},
    Publisher= {Elsevier},
}

@article{smyth2003pearson,
  title={Pearson's goodness of fit statistic as a score test statistic},
  author={Smyth, Gordon K},
  journal={Lecture notes-monograph series},
  pages={115--126},
  year={2003},
  publisher={JSTOR}
}

@book{bickel1993efficient,
  title={Efficient and adaptive estimation for semiparametric models},
  author={Bickel, Peter J and Klaassen, Chris AJ and Ritov, Ya’acov and Wellner, Jon A},
  volume={4},
  year={1993},
  publisher={Johns Hopkins University Press Baltimore}
}

@book{lindsey1996parametric,
  title={Parametric statistical inference},
  author={Lindsey, James K},
  year={1996},
  publisher={Oxford University Press}
}

@article{waterman1996,
  title={A simple and accurate method for approximate conditional inference applied to exponential family models},
  author={Waterman, Richard P and Lindsay, Bruce G},
  journal={Journal of the Royal Statistical Society: Series B (Methodological)},
  volume={58},
  number={1},
  pages={177--188},
  year={1996},
  publisher={Wiley Online Library}
}

@article{cox1987parameter,
  title={Parameter orthogonality and approximate conditional inference},
  author={Cox, David Roxbee and Reid, Nancy},
  journal={Journal of the Royal Statistical Society: Series B (Methodological)},
  volume={49},
  number={1},
  pages={1--18},
  year={1987},
  publisher={Wiley Online Library}
}

@article{barndorff-nielsen_approximate_1990,
	title = {Approximate {Interval} {Probabilities}},
	volume = {52},
	number = {3},
	journal = {Journal of the Royal Statistical Society. Series B (Methodological)},
	author = {Barndorff-Nielsen, O. E.},
	year = {1990},
	pages = {485--496},
}

@article{price_principal_2006,
	title = {Principal components analysis corrects for stratification in genome-wide association studies},
	volume = {38},
	number = {8},
	journal = {Nature Genetics},
	author = {Price, Alkes L. and Patterson, Nick J. and Plenge, Robert M. and Weinblatt, Michael E. and Shadick, Nancy A. and Reich, David},
	month = aug,
	year = {2006},
	pages = {904--909},
}

@article{patterson_population_2006,
	title = {Population {Structure} and {Eigenanalysis}},
	volume = {2},
	number = {12},
	journal = {PLOS Genetics},
	author = {Patterson, Nick and Price, Alkes L. and Reich, David},
	year = {2006},
	pages = {e190},
}

@article{skovgaard1987saddlepoint,
  title={Saddlepoint expansions for conditional distributions},
  author={Skovgaard, Ib M},
  journal={Journal of Applied Probability},
  volume={24},
  number={4},
  pages={875--887},
  year={1987},
  publisher={Cambridge University Press}
}

@article{jannot2015p,
  title={{$P< 5\times 10^{- 8}$} has emerged as a standard of statistical significance for genome-wide association studies},
  author={Jannot, Anne-Sophie and Ehret, Georg and Perneger, Thomas},
  journal={Journal of clinical epidemiology},
  volume={68},
  number={4},
  pages={460--465},
  year={2015},
  publisher={Elsevier}
}

@article{lugannani_rice_1980, title={Saddle point approximation for the distribution of the sum of independent random variables}, volume={12}, DOI={10.2307/1426607}, number={2}, journal={Advances in Applied Probability}, publisher={Cambridge University Press}, author={Lugannani, Robert and Rice, Stephen}, year={1980}, pages={475–490}}

@article{LRfails,
	title = {An example in which the {Lugannani}-{Rice} saddlepoint formula fails},
	volume = {23},
	number = {1},
	journal = {Statistics \& Probability Letters},
	author = {Booth, James G. and Wood, Andrew T. A.},
	month = apr,
	year = {1995},
	pages = {53--61},
}

@article{liu_acat_2019,
	title = {{ACAT}: {A} {Fast} and {Powerful} p {Value} {Combination} {Method} for {Rare}-{Variant} {Analysis} in {Sequencing} {Studies}},
	volume = {104},
	number = {3},
	journal = {The American Journal of Human Genetics},
	author = {Liu, Yaowu and Chen, Sixing and Li, Zilin and Morrison, Alanna C. and Boerwinkle, Eric and Lin, Xihong},
	month = mar,
	year = {2019},
	pages = {410--421},
}

@article{skat,
	title = {Rare-{Variant} {Association} {Testing} for {Sequencing} {Data} with the {Sequence} {Kernel} {Association} {Test}},
	volume = {89},
	number = {1},
	journal = {American Journal of Human Genetics},
	author = {Wu, Michael C. and Lee, Seunggeun and Cai, Tianxi and Li, Yun and Boehnke, Michael and Lin, Xihong},
	month = jul,
	year = {2011},
	pages = {82--93},
}

@techreport{van_hout_whole_2019,
	title = {Whole exome sequencing and characterization of coding variation in 49,960 individuals in the {UK} {Biobank}},
	author = {Van Hout, Cristopher V. and Tachmazidou, Ioanna and Backman, Joshua D. and Hoffman, Joshua X. and Ye, Bin and Pandey, Ashutosh K. and Gonzaga-Jauregui, Claudia and Khalid, Shareef and Liu, Daren and Banerjee, Nilanjana and Li, Alexander H. and Colm, O’Dushlaine and Marcketta, Anthony and Staples, Jeffrey and Schurmann, Claudia and Hawes, Alicia and Maxwell, Evan and Barnard, Leland and Lopez, Alexander and Penn, John and Habegger, Lukas and Blumenfeld, Andrew L. and Yadav, Ashish and Praveen, Kavita and Jones, Marcus and Salerno, William J. and Chung, Wendy K. and Surakka, Ida and Willer, Cristen J. and Hveem, Kristian and Leader, Joseph B. and Carey, David J. and Ledbetter, David H. and {Geisinger-Regeneron DiscovEHR Collaboration} and Cardon, Lon and Yancopoulos, George D. and Economides, Aris and Coppola, Giovanni and Shuldiner, Alan R. and Balasubramanian, Suganthi and Cantor, Michael and Nelson, Matthew R. and Whittaker, John and Reid, Jeffrey G. and Marchini, Jonathan and Overton, John D. and Scott, Robert A. and Abecasis, Gonçalo and Yerges-Armstrong, Laura and Baras, Aris and {on behalf of the Regeneron Genetics Center}},
	month = mar,
	year = {2019},
}

\end{document}


\maketitle

\section{Quality assessment of UK Biobank Genetic Data}\label{QCAppendix}

\subsection{Common variants}

Analyses were limited to autosomal variants covered by both genotype arrays used over the course of the study and that passed the batch-level quality control. SNPs were included if the call rate was above 99\%, the Hardy-Weinberg equilibrium $p$-value was less than $5 \cdot 10^{-8}$, and the minor allele frequency was larger than 1\%. 529,024 SNPs passed these filters.

Individuals were removed if the genetic and reported sex did not match and if the sex chromosomes were not XX or XY. Outliers in heterozygosity and missing rates were removed. The analyses were limited to those identified as Caucasian through the UK Biobank’s PCA analysis (field 22006). All individuals had an individual call rate larger than 99\%. 366,752 individuals passed these filters. Individuals were removed if the genetic and reported sex did not match and if the sex chromosomes were not XX or XY.

\subsection{Rare variants}
Analyses were limited to autosomal variants at chromosome 6. Details of quality assessment of the sequenced exomes from 49,960 UKB participants is given in \cite{van_hout_whole_2019}. For further quality assessment, out of these 49,960 participants, the analysis were limited to those identified as Caucasian through the UK Biobank’s PCA analysis (field 22006). Individuals were removed if the genetic and reported sex did not match and if the sex chromosomes were not XX or XY. The SNPs had a MAF less than 0.01, but a MAC larger than two. All SNPs had a missing rate less than 0.01, and all individuals had an individual call rate larger than 99\%.

\section{Approximating the double saddlepoint method by a normal approximation to $\bm U_{\beta}$}

By a double saddlepoint method we may, under the null, estimate 
\begin{align}
P( U_{\gamma} = u_{\gamma} | \bm U_{\bm \beta} = \bm 0) = \frac{f(\bm 0, u_{\gamma})}{f_{\bm \beta}(\bm 0)},
\label{eq:cond_density}
\end{align}
by using saddlepoint techniques to approximate the joint distribution $f$ of $\bm U_{\bm \beta}$ and $U_{\gamma}$ at $\bm U_{\bm \beta} = \bm 0$, and the marginal distribution $f_{\bm \beta}$ of $\bm U_{\bm \beta}$ at $\bm U_{\bm \beta} = \bm 0$.

\subsection{The joint cumulant generating function}

Let $\bm t$ be a vector of dimension $d+1$, which we partition into the vector $\bm t_{\beta}$ of dimension $d$ and the scalar $t_{\gamma}$. Since $Y_i \sim \text{binomial}(\mu_i)$, the CGF of $\bm U$ may then be expressed as
\begin{align}
K(\bm t) = K(\bm t_{\bm \beta}, t_{\gamma})
 & = \sum_{i=1}^n \ln \left( 1 - \mu_i + \mu_i \exp(g_i t_{\gamma} + \bm t_{\bm \beta}^T \bm x_{i})  \right) \nonumber \\ & - t_{\gamma} \sum_{i=1}^n g_i \mu_i - \bm t_{\bm \beta}^T \sum_{i=1}^n \bm x_{i} \mu_i.
 \label{eq:cgf_joint}
\end{align}
Derivatives of $K(\bm t_{\beta}, t_{\gamma})$ with respect to $\bm t_{\beta}$ and $t_{\gamma}$, denoted $\nabla_{\bm t_{\bm \beta}} K(\bm t_{\beta}, t_{\gamma})$ and $\frac{\partial}{\partial t_\gamma} K(\bm t_{\beta}, t_{\gamma})$ respectively, are
\begin{align*}
& \nabla_{\bm t_{\bm \beta}} K(\bm t_{\bm \beta}, t_{\gamma}) = \sum_{i=1}^n \mu_i \left(\frac{\exp(g_i t_{\gamma} + \bm t_{\bm \beta}^T \bm x_{i}) }{\left( 1 - \mu_i + \mu_i \exp(g_i t_{\gamma} + \bm t_{\bm \beta}^T \bm x_{i}) \right)} - 1\right) \bm x_{i},
\end{align*}
and
\begin{align*}
& \frac{\partial}{\partial t_\gamma} K(\bm t_{\bm \beta}, t_{\gamma}) = \sum_{i=1}^n  \mu_i\left(\frac{\exp(g_i t_{\gamma} + \bm t_{\bm \beta}^T \bm x_{i}) }{\left( 1 - \mu_i + \mu_i \exp(g_i t_{\gamma} + \bm t_{\bm \beta}^T \bm x_{i}) \right)} - 1\right)g_{i},
\end{align*}
so that the gradient of $K(\bm t_{\bm \beta}, t_{\gamma})$, denoted $\nabla K(\bm t_{\bm \beta}, t_{\gamma})$, may be expressed as
\begin{align*}
\nabla K(\bm t_{\bm \beta}, t_{\gamma}) = \begin{pmatrix}
\nabla_{\bm t_{\bm \beta}} K(\bm t_{\bm \beta}, t_{\gamma}) \\
\frac{\partial}{\partial t_\gamma} K(\bm t_{\bm \beta}, t_{\gamma})
\end{pmatrix}.
\end{align*}
Let $\bm \theta = \smash{\begin{pmatrix}\bm \beta^\text T& \gamma\end{pmatrix}}^\text T$ denote the full parameter set, and define a diagonal matrix $M^{\bm \theta}$ with entries 
\begin{align*}
M^{\bm \theta}_{ii} = \frac{\mu_i(1-\mu_i)  \exp(- g_i t_{\gamma} - \bm t_{\bm \beta}^T \bm x_{i})}{\left( (1-\mu_i)\exp(- g_i t_{\gamma} - \bm t_{\bm \beta}^T \bm x_{i}) + \mu_i  \right)^2}.
\end{align*}
The Hessian of $K$, denoted $H(\bm t)$, can be expressed as 
\begin{align*}
H(\bm t) 
= 
\begin{bmatrix}
\frac{\partial^2}{\partial \bm \beta \partial \bm \beta^T}K(\bm t_{\bm \beta}, t_{\gamma}) & \frac{\partial^2}{\partial \bm \beta \partial t_\gamma}K(\bm t_{\bm \beta}, t_{\gamma}) \\ \frac{\partial^2}{\partial t_\gamma \partial \bm \beta}K(\bm t_{\bm \beta}, t_{\gamma})& 
\frac{\partial^2}{\partial t_\gamma^2}K(\bm t_{\bm \beta}, t_{\gamma})
\end{bmatrix} 
= 
\begin{bmatrix}
X^T M \bm X & X^T M \bm g \\  \bm g^T M X & 
\bm g^T M \bm g
\end{bmatrix}.
\end{align*}

\subsection{The marginal cumulant generating function}
The cumulant generating function of $\bm U_{\bm \beta}$, denoted $K_{\bm \beta}(\bm t_{\bm \beta})$, is given by
\begin{align}
K_{\bm \beta}(\bm t_{\bm \beta})
 & = \sum_{i=1}^n \ln \left( 1 - \mu_i + \mu_i \exp\left(\bm t_{\bm \beta}^T \bm x_{i}\right) \right)  -\sum_{i=1}^n \bm t_{\bm \beta}^T\bm x_{i}\mu_i  .
 \label{eq:cgf_marginal}
\end{align}
The gradient, denoted $\nabla K_{\bm \beta}(\bm t_{\bm \beta})$,  is 
\begin{align*}
\nabla K_{\bm \beta}(\bm t_{\bm \beta}) = \sum_{i=1}^n \mu_i  \left(\frac{ \exp( \bm t_{\bm \beta}^T \bm x_{i}) }{\left( 1 - \mu_i + \mu_i \exp(g_i t_{\gamma} + \bm t_{\bm \beta}^T \bm x_{i}) \right)} - 1\right) \bm x_{i},
\end{align*}
and the Hessian, denoted $H_{\bm \beta}(\bm t_{\bm \beta})$, is
\begin{align*}
H_{\bm \beta}(\bm t_{\bm \beta}) = X^T M^{\bm \beta} X,
\end{align*}
where $M^{\bm \beta}$ is a diagonal matrix with entries 

\[M^{\bm \beta}_{ii} = \frac{\mu_i(1-\mu_i)  \exp( - \bm t_{\bm \beta}^T \bm x_{i} )}{\left( (1-\mu_i)\exp(- \bm t_{\bm \beta}^T \bm x_{i}) + \mu_i  \right)^2}.\]
We note that
\begin{align}
K(\bm t_{\bm \beta}, 0) = K_{\bm \beta}(\bm t_{\bm \beta}).
\label{eq:equalities_marginal}
\end{align}

\subsection{Double-saddlepoint approximation}

The saddlepoint approximation of the probability distribution of the score vector $\bm U$ evaluated at $\bm U_{\bm \beta}= 0$ is given by
\begin{align*}
\hat{f}(\bm 0, u_{\gamma}) = (2\pi)^{-(d+1)/2} |H(\hat{\bm t})|^{-1/2} \exp\left\lbrace  K(\hat{\bm t}_{\bm \beta}, \hat{t}_{\gamma}) - \hat{t}_{\gamma}u_{\gamma}) \right\rbrace,
\end{align*}
where $\smash{\begin{pmatrix}\hat{\bm t}_{\bm \beta}^\text T&\hat{t}_{\gamma}\end{pmatrix}}^\text T$ is the $(d+1)$-dimensional saddlepoint that solves $K'(\hat{\bm t}_{\bm \beta}, \hat{t}_{\gamma}) = \smash{\begin{pmatrix}\bm 0^\text T &u_{\gamma} \end{pmatrix}}^\text T$. The saddlepoint approximation of the marginal distribution of $\bm U_{\bm \beta}$, evaluated at $\bm U_{\bm \beta} = \bm 0$ is similarly
\begin{align*}
\hat{f}_{\bm \beta}(\bm 0) = (2\pi)^{-d/2} |H_{\bm \beta}(\tilde{\bm t}_{\bm \beta})|^{-1/2} \exp\left\lbrace  K_{\bm \beta}(\tilde{\bm t}_{\bm \beta}) \right\rbrace,
\end{align*}
where $\tilde{\bm t}_{\bm \beta}$ is the $d$-dimensional saddlepoint that solves $\nabla K_{\bm \beta}(\bm t_{\bm \beta}) = \bm 0$. We showed in Appendix B that $\tilde{\bm t}_{\bm \beta} = \bm 0$, hence 
\begin{align*}
\hat{f}_{\bm \beta}(\bm 0) = (2\pi)^{-d/2} |H_{\bm \beta}(\bm 0)|^{-1/2} \exp\left\lbrace  K_{\bm \beta}(\bm 0) \right\rbrace.
\end{align*}

\subsection{Speed-up algorithm}

Starting with the joint CGF $K(\bm t_{\bm \beta}, t_{\gamma})$ in Equation \eqref{eq:cgf_joint} we split the sum into the sets of individuals with and without minor alleles:
\begin{align*}
K(\bm t_{\bm \beta}, t_{\gamma}) 
& = \sum_{i=1}^m \ln \left( 1 - \mu_i + \mu_i \exp(g_i t_{\gamma} + \bm t_{\bm \beta}^T \bm x_{i}) \right) - t_{\gamma} \sum_{i=1}^m g_i \mu_i - \bm t_{\bm \beta}^T \sum_{i=1}^m \bm x_{i} \mu_i  \\
& + \sum_{i=m+1}^n \ln \left( 1 - \mu_i + \mu_i \exp( \bm t_{\bm \beta}^T \bm x_{i}) \right) - \bm t_{\bm \beta}^T \sum_{i=m+1}^n \bm x_{i} \mu_i.
\end{align*}

By comparing with Equation \eqref{eq:cgf_marginal}, the last two terms is in fact the part of the cumulant generating function of $\bm U_{\beta}$ restricted to individuals with $g_i = 0$, denoted $K_{\bm \beta}^*(\bm t_{\bm \beta})$. As discussed in \cite{dey_fast_2017}, if the non-genetic covariates are not particularly skewed, then a normal approximation to $\bm U_{\bm \beta}$ may be accurate. If there are few individuals with $g_i >0$, which is typically the case, this would imply that a normal approximation of $\bm U^{*}_{\bm \beta}$, the part of $\bm U_{\bm \beta}$ with $g_i=0$, may also be accurate. Therefore, let $X_{g_i=0}$, $\bm Y_{g_i=0}$ and $\bm \mu_{g_i=0}$ denote the part of $X$, $\bm Y$ and $\bm \mu$ with $g_i = 0$, and so $\bm U^{*}_{\bm \beta} = X_{g_i=0}^T(\bm Y_{g_i=0} - \bm \mu_{g_i=0})$ with $E(\bm U^{*}_{\bm \beta}) = \bm 0$ and $\Cov(\bm U^{*}_{\bm \beta}) = X_{g_i=0}^T W_{g_i=0}  X_{g_i=0}$, with $W_{g_i=0}$ the submatrix of the diagonal matrix $W$ with entries $\mu_i(1-\mu_i)$ among those individuals with $g_i = 0$. By approximating $\bm U^{*}_{\bm \beta}$ to have a normal distribution, the approximation of the CGF of $U^{*}_{\bm \beta}$ is  $K_{\bm \beta}^*(\bm t_{\bm \beta}) \approx \frac{1}{2} \bm t_{\bm \beta}^T \Cov(\bm U^{*}_{\bm \beta}) \bm t_{\bm \beta}$. And consequently, the original CGF may be approximated as
\begin{align*}
K(\bm t_{\bm \beta}, t_{\gamma})  & \approx \sum_{i=1}^m \ln \left( 1 - \mu_i + \mu_i \exp(g_i t_{\gamma} + \bm t_{\bm \beta}^T \bm x_{i}) \right) - t_{\gamma} \sum_{i=1}^m g_i \mu_i - \bm t_{\bm \beta}^T \sum_{i=1}^m \bm x_{i}^T \mu_i  \\ & +  \frac{1}{2} \bm t_{\bm \beta}^T \Cov(\bm U^{*}_{\bm \beta}) \bm t_{\bm \beta}.
\end{align*}
This approximation to the CGF does not represent any reasonable speed-up yet, since $\Cov(\bm U^{*}_{\bm \beta})$ must be computed for each genetic variant and requires $O(n-m)$ calculations. However, we may express for each variant $\Cov(\bm U^{*}_{\bm \beta}) = \Cov(\bm U_{\bm \beta})  - \Cov(\bm U^{\cross[0.4pt]}_{\bm \beta})$, with $\Cov(\bm U_{\bm \beta})$ the same for all variants, while $\bm U^{\cross[0.4pt]}_{\bm \beta}$ is the part of $\bm U_{\bm \beta}$ with $g_i > 0$, and so $\Cov(\bm U^{\cross[0.4pt]}_{\bm \beta}) = X_{g_i>0}^T W_{g_i>0}  X_{g_i>0}$. As $\Cov(\bm U_{\bm \beta})$ can be precomputed for all variants, this requires $O(m)$ calculations. Hence, the approximation
\begin{align*}
K(\bm t_{\bm \beta}, t_{\gamma})  & \approx \sum_{i=1}^m \ln \left( 1 - \mu_i + \mu_i \exp(g_i t_{\gamma} + \bm t_{\bm \beta}^T \bm x_{i}) \right) - t_{\gamma} \sum_{i=1}^m g_i \mu_i - \bm t_{\bm \beta}^T \sum_{i=1}^m \bm x_{i}^T \mu_i  \\ & +  \frac{1}{2} \bm t_{\bm \beta}^T \left( \Cov(\bm U_{\bm \beta})  - \Cov(\bm U^{\cross[0.4pt]}_{\bm \beta})\right) \bm t_{\bm \beta}
\end{align*}
is computed only for those individuals with $g_i > 0$ which leads to a substantial reduction in running time from $O(n-m)$ to $O(m)$ calculations when $m << n$, which is typically the case, and particularly relevant for rare variants.

Similarly as for the normal approximation of $K(\bm t_{\bm \beta}, t_{\gamma})$, the normal approximation of $K_{\bm \beta}(\bm t_{\bm \beta})$ is given by
\begin{align*}
K_{\bm \beta}(\bm t_{\bm \beta}) \approx \frac{1}{2} \bm t_{\bm \beta}^T \Cov(\bm U_{\bm \beta})  \bm t_{\bm \beta}.
\end{align*}


\section{Simulations of genetic association studies with an imbalanced response}

We will in detail explain the simulations given in Section 5.3. The simulation set-up is motivated by \cite{dey_fast_2017} by conditioning on the which individuals are cases and controls, say $\bm y = (\bm 0_c,\bm 0_{n-c}^{T})$, with $\bm 0_c$ and $\bm 0_{n-c}$ vectors of zeros with size $c$ (the number of cases) and $n-c$ (the number of controls). We consider the logistic regression model
\begin{align*}
\text{logit}(\mu_i) = \beta_0 + x_{i,1} + x_{i,2} + \gamma g_i,
\end{align*}
with $X_{1}\sim \text{Bernoulli}(0.5)$, $X_{2} \sim N(0,1)$ and $G \sim \text{binom}(2,\text{MAF})$ mutually independent. Since we are evaluating validity of tests, we set $\gamma = 0$. For each iteration, the genotype vector is sampled independently of $\bm y$, while the nuisance covariates for each individual are sampled conditionally on $\bm y$. In each iteration, and for each method, we record whether a false rejection has occurred. With a total of $10^9$ iterations, we estimate the type I error rate conditioned on the constant phenotype vector $\bm y$ when applying SPA, ESPA-CC and DSPA-CC.  We want $\beta_0$ to be such that the disease prevalence to be $1 \%$ in the population, i.e. $P(Y=1) = 0.01$. That is we want:

\begin{align}
    \begin{split}
        &P(Y = 1) = \int_{x_{1}=-\infty}^{\infty} \sum_{x_{1}=0}^1 P(Y = 1 | x_{1},x_{2}) P(x_{1},x_{2}) dx_{1} \\
        &=P(Y = 1) = \int_{x_{1}=-\infty}^{\infty} \sum_{x_{1}=0}^1 P(Y = 1 | x_{1},x_{2}) P(x_{1})P(x_{2}) dx_{1}\\
        &= 0.5\sqrt{\frac{1}{2 \pi}} \cdot \int_{x_2=-\infty}^{\infty}\exp(-0.5x^2_2)\left(\frac{1}{1+ \exp(-\beta_0-x_{2})}  + \frac{1}{1+ \exp(-\beta_0-1-x_{2})}\right)dx_2 = 0.01
    \end{split}
\end{align}

A solution is $\beta_0 = -5.6$.\\

Given the vector of phenotype values $\bm y$, the nuisance covariates need to be sampled according to their conditional probabilities:

\begin{align}
    \begin{split}
        &P(x_1|y) = P(x_1,y)/P(y) = P(y|x_1)P(x_1)/P(y) \\
        &= P(x_1)/P(y_1)\int_{x_2=-\infty}^{\infty}  P(y|x_1,x_2)P(x_2) dx_2,
    \label{x1condy}
    \end{split}
\end{align}

and

\begin{align}
    \begin{split}
        &P(x_2|y) = P(x_2,y)/P(y) = P(y|x_2)P(x_2)/P(y) \\
        &= P(x_2)/P(y_1) \sum_{x_1=0}^1 P(y|x_1,x_2)P(x_2),
    \label{x2condy}
    \end{split}
\end{align}

Therefore from \eqref{x1condy} (with prevalence 0.01):

\begin{equation}
    P(X_1 =x_1|y=1) = 50\sqrt{\frac{1}{2 \pi}} \cdot \int_{x_2 = -\infty}^{\infty} \frac{\exp(-0.5x^2)}{1+ \exp(-\beta_0-x_1-x_{2})} dx_2,
\end{equation}

and similarly one can compute $P(X_1 =x_1|y=0)$ for each value of $x_1$. For the sampling of $x_2$ conditional on $y$ we will get for instance:

\begin{align}
    \begin{split}
        &P(x_2|y=1) = 50 \sqrt{\frac{1}{2 \pi}}\exp(-0.5x^2) \left(\frac{1}{1+ \exp(-\beta_0-x_{2})} + \frac{1}{1+ \exp(-\beta_0-1-x_{2})}\right).
    \end{split}
\end{align}

The continuous probability distribution of $P(x_2|y=1)$, as well as for $P(x_2|y=0)$, does not belong to any known distribution class, however one can see that $P(x_2|y=1) \le \phi(x_2)$ for all $x_2$ with $\phi()$ the standard normal distribution. Therefore the standard normal can be used as a proposal distribution in a \textit{rejection sampling} procedure. However, the large amount of simulations needed requires a faster approach as the efficiency in the rejection sampling depends on how close the proposal distribution resembles the true distribution. It can be shown in this particular case that the efficiency decreases for decreasing prevalence ($P(Y=1)$). Since the probability distribution of $P(x_2|y=1)$ as well as $P(x_2|y=0)$ can be shown to be log-concave, one can apply the much more efficient \textit{adaptive rejection sampling} procedure, where the proposal distribution is adaptively improved during the iterations.

\section{PCA plots}

\begin{figure}[h]
\centering
\includegraphics[width=0.65\textwidth]{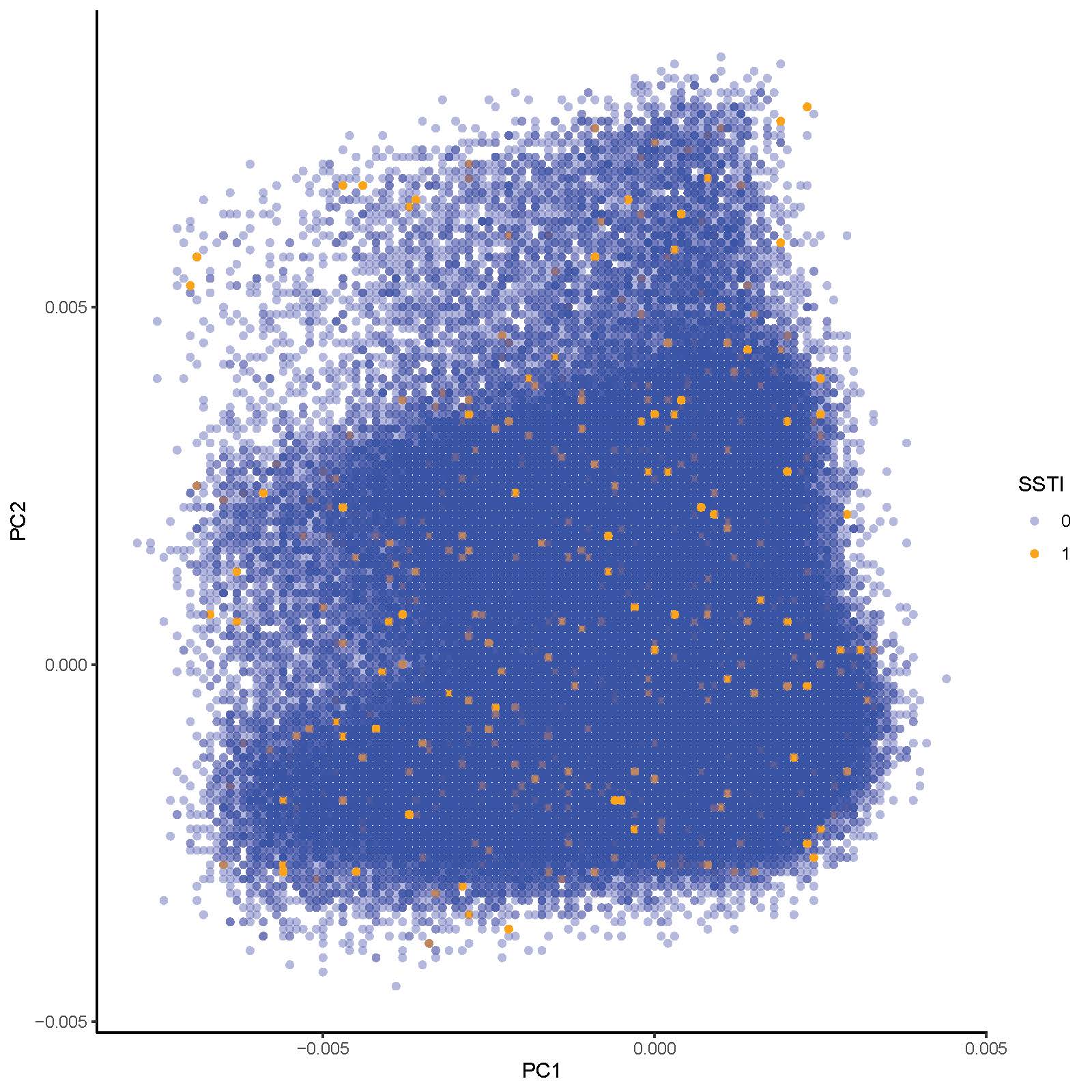}
\caption{PCA plot of first and second principal components when analysing the common variants.}
\label{TIEUB}
\end{figure}

\begin{figure}[t!]
\centering
\includegraphics[width=0.6\textwidth]{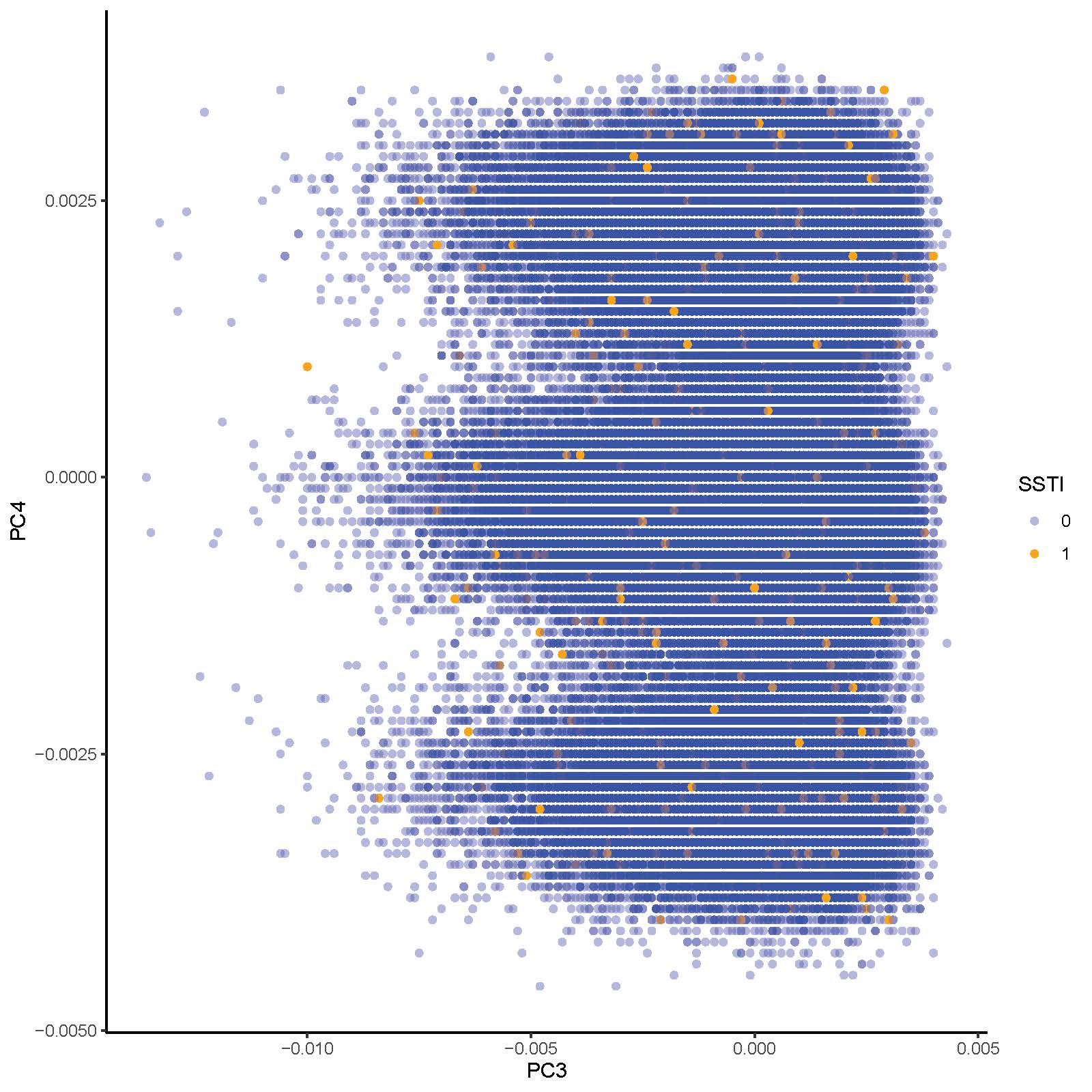}
\caption{PCA plot of third and fourth principal components when analysing the common variants.}
\label{TIEUB2}
\end{figure}

\begin{figure}[b!]
\centering
\includegraphics[width=0.6\textwidth]{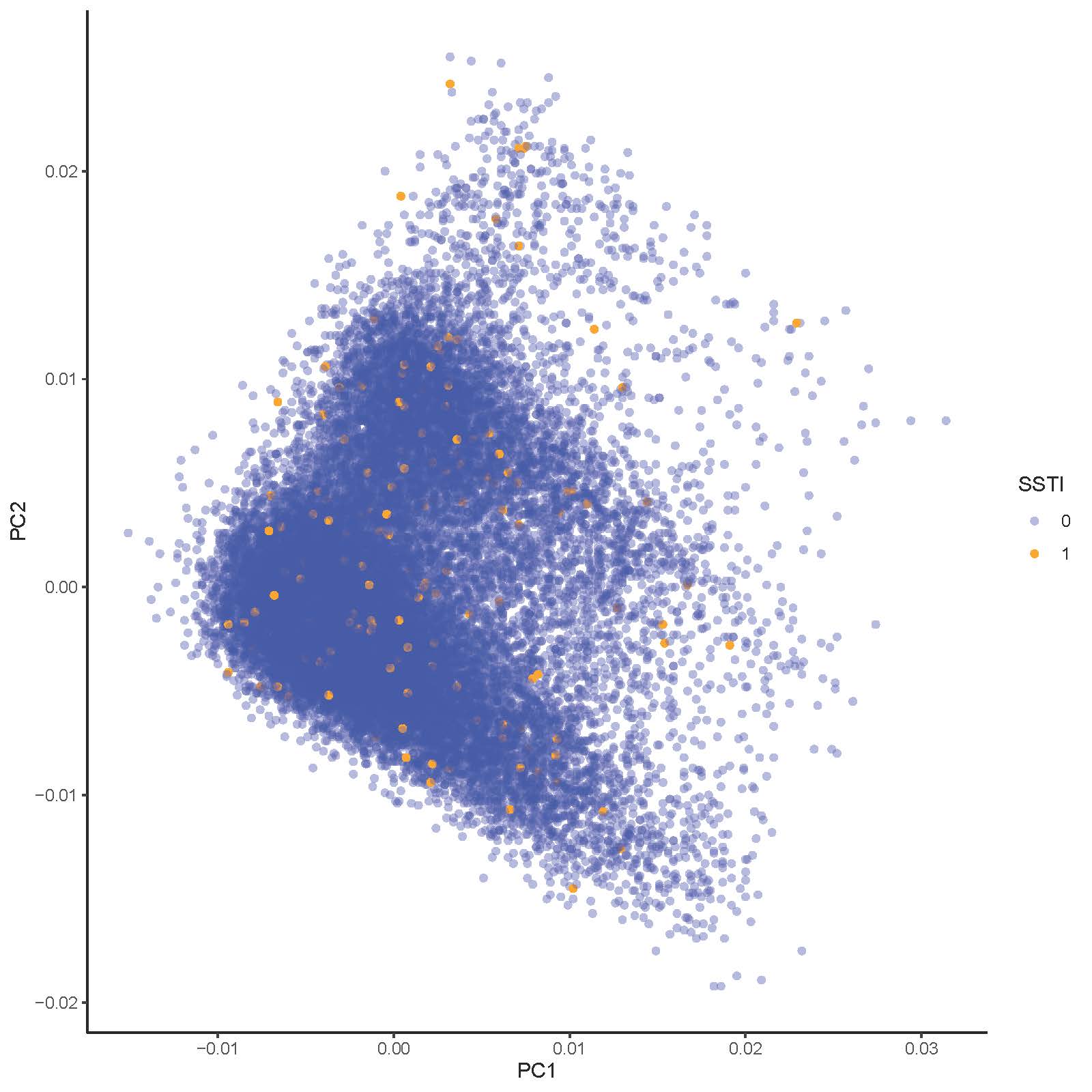}
\caption{PCA plot of first and second principal components when analysing the rare variants.}
\label{TIEUB3}
\end{figure}

\clearpage 

\begin{figure}[t!]
\centering
\includegraphics[width=0.65\textwidth]{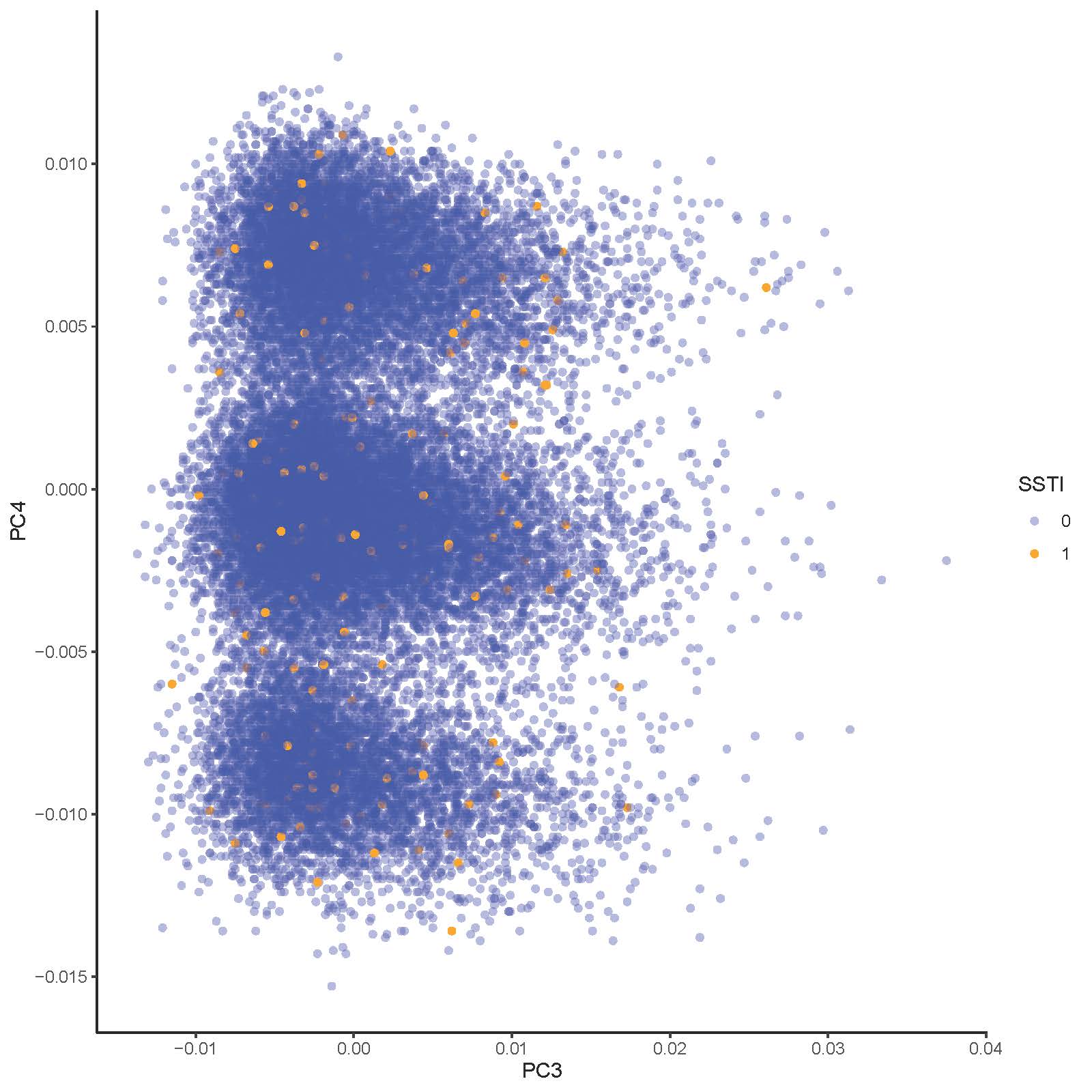}
\caption{PCA plot of first and second principal components when analysing the rare variants.}
\label{TIEUB4}
\end{figure}
\printbibliography